\documentclass[aps,prd,preprint,superscriptaddress,showpacs,ctexart]{revtex4-1} 
\pdfoutput=1 
\usepackage{float}
\usepackage[bottom]{footmisc}
\usepackage{caption}
\usepackage{textcomp}
\usepackage{amsmath}
\usepackage{amssymb}
\usepackage{graphicx}
\usepackage{subfigure}
\usepackage{diagbox}
\usepackage{color}
\usepackage{ulem}
\usepackage{setspace}
\usepackage[unicode=true, bookmarks=false, breaklinks=false,pdfborder={0 0 1},colorlinks=false] {hyperref}

\makeatletter

\newcommand{\bmat}{\left(\begin{array}}
\newcommand{\emat}{\end{array}\right)}
\newcommand{\beq}{\begin{equation}}
\newcommand{\eeq}{\end{equation}}

\makeatletter 
\@addtoreset{equation}{section}
\makeatother  

\def\alt{\mathrel{\mathpalette\gl@align<}}
\def\agt{\mathrel{\mathpalette\gl@align>}}
\def\gl@align#1#2{\lower.6ex\vbox{\baselineskip\z@skip\lineskip\z@
\ialign{$\m@th#1\hfil##\hfil$\crcr#2\crcr\sim\crcr}}}
\def\su5u1{SU(5) \times U(1)}
\def\fsu5u1{SU(5) \times U(1)'}
\def\so10{SO(10)}
\def\sq20{SO(10) \times SO(10)}

\def\bwt{\begin{widetext}}
\def\ewt{\end{widetext}}
\def\be{\begin{equation}}
\def\ee{\end{equation}}
\def\bea{\begin{eqnarray}}
\def\eea{\end{eqnarray}}
\def\bean{\begin{eqnarray*}}
\def\eean{\end{eqnarray*}}
\def\bary{\begin{array}}
\def\eary{\end{array}}
\def\bit{\begin{itemize}}
\def\eit{\end{itemize}}

\makeatother

\begin{document}

\begin{center}

{\Large \bf S-duality and loop operators in canonical formalism   \\}

\end{center}

\vspace{7 mm}

\begin{center}

{ Shan Hu }

\vspace{6mm}
{\small \sl Department of Physics, Faculty of Physics and Electronic Sciences, Hubei University,} \\
\vspace{3mm}
{\small  \sl Wuhan 430062, People’s Republic of China} \\

\vspace{6mm}

{\small \tt hushan@hubu.edu.cn} \\

\end{center}

\vspace{8 mm}

\begin{abstract}\vspace{1cm}

We study the gauge invariant 't Hooft operator in canonical formalism for Yang-Mills theory as well as the $\mathcal{N} =4 $ super-Yang-Mills theory with the gauge group $ U(N) $. It is shown that the spectrum of the 't Hooft operator labeled by the arbitrary irreducible representation of the gauge group is the same as the spectrum of the Wilson operator labeled by the same representation. So it is possible to construct a unitary operator $ S $ making the two kinds of loop operators transformed into each other. S-duality transformation could be realized by the operator $ S $. We compute the supersymmetry variations of the loop operators with the fermionic couplings turned off. The result is consistent with the expectation that the action of $ S $ should make supercharges transform with a $ U(1)_{Y} $ phase.

\end{abstract}

\maketitle

\begin{spacing}{1.8}
\tableofcontents
\end{spacing}

\section{Introduction}

It is well known that the source-free Maxwell theory exhibits the electric-magnetic duality (S-duality). The duality can also be extended to nonlinear electrodynamics such as the Born-Infeld theory describing the dynamics of a $ D3 $ brane \cite{1y,2y,3y,980y}. Generically, for $ U(1) $ gauge theory with the coupling constant $ g^{2} $ and the Lagrangian $ L(g^{2};F_{\mu\nu}) $, the dual field strength can be obtained by adding a Lagrange multiplier (dual potential) \cite{LM}: 
\begin{equation}\label{ml11}
	L'(g^{2};F_{\mu\nu},\tilde{A}_{\mu})=L(g^{2};F_{\mu\nu})-\tilde{A}_{\mu}G^{\mu}\;\;\;\;\;\;\;\;\;G^{\mu}=\frac{1}{2}\epsilon^{\mu\nu\rho\sigma}\partial_{\nu}F_{\rho\sigma}\;.
\end{equation}
The saddle-point equations are
\begin{equation}\label{ml1}
	\frac{\delta L'}{\delta \tilde{A}_{\mu}}=-G^{\mu}=-\frac{1}{2}\epsilon^{\mu\nu\rho\sigma}\partial_{\nu}F_{\rho\sigma}=0
\end{equation}
and 
\begin{equation}\label{ml2}
		\frac{\delta L'}{\delta F_{\mu\nu}}=\frac{\delta L}{\delta F_{\mu\nu}}+\frac{1}{2}\epsilon^{\rho\sigma\mu\nu}\partial_{\sigma}\tilde{A}_{\rho}=0\;.
\end{equation}
Equation (\ref{ml1}) is the Bianchi identity, while (\ref{ml2}) implies the equations of motion $\partial_{\mu}(\delta L/\delta F_{\mu\nu})=0$. The dual field strength is 
\begin{equation}\label{12q}
\tilde{F}_{\mu\nu}=-\epsilon_{\mu\nu\rho\sigma} \delta L/\delta F_{\rho\sigma}= \partial_{\mu} \tilde{A}_{\nu}-\partial_{\nu} \tilde{A}_{\mu}
\end{equation}
with $ \tilde{A}_{\mu} $ the dual potential. The Bianchi identity of $\tilde{F}_{\mu\nu}$ gives the equations of motion for $F_{\mu\nu}$ and vice versa. If the equations of motion for $F_{\mu\nu}  $ and $  \tilde{F}_{\mu\nu}$ take the same form except for the replacement of $ g^{2} $ by $ 1/g^{2} $, the theory will be duality invariant. This is the situation for Maxwell theory and Born-Infeld theory.

In non-Abelian theories, (\ref{ml11}) can be replace by  \cite{LM}
\begin{equation}
	L'(g;F_{\mu\nu},\tilde{A}_{\mu})=L(g;F_{\mu\nu})-tr(\tilde{A}_{\mu}G^{\mu})\;,\;\;\;\;\;\;\;\;G^{\mu}=\frac{1}{2}\epsilon^{\mu\nu\rho\sigma}D_{\nu}F_{\rho\sigma}\;. 
\end{equation}
However, $ \tilde{A}_{\mu}   $ transforms as $ \tilde{A}_{\mu}\rightarrow u\tilde{A}_{\mu}u^{-1} $ under the local gauge transformation thus is not a gauge potential in the ordinary sense. For other possibilities, $ \tilde{F}_{\mu\nu} =-\epsilon_{\mu\nu\rho\sigma} \delta L/\delta F_{\rho\sigma}  $ is not the field strength, from which, some particular gauge potential can be constructed, see remarks around (8.4) in \cite{LM}.

Nevertheless, S-duality indeed has the non-Abelian generalizations \cite{1,2,3,3gf,3gff}. Especially, $ \mathcal{N} =4$ super-Yang-Mills (SYM) theory with the gauge group $ U(N) $ and the coupling constant $ \tau $ is dual to the same theory with the coupling constant $ -1/\tau $. Explicit conjectures have been made for the S-duality actions on local operators \cite{456, op3, op2}, line operators \cite{10, wh2}, surface operators \cite{14, 15}, and domain walls \cite{16, 17}. Compared with the Abelian case, our understanding for the non-Abelian S-duality transformation rule is still limited. Given an arbitrary gauge invariant operator or state, there is no systematic way to determine its S-dual.

The duality transformation in (\ref{ml11})-(\ref{12q}) is Lagrangian dependent, which may increase the complexity. $ D3 $ branes come from $ M5 $ branes wrapping on $ T^{2} $, with the action of S-duality realized as the $ SL(2,Z) $ transformation of $ T^{2} $. So, similar with the rotation in the $ 3d $ external space, S-duality transformation should also have a kinematical formulation.

The dynamical and the kinematical information can be neatly disentangled in canonical formalism. For $ U(1) $ gauge theory, the canonical coordinates are $ A_{i} $ with the conjugate momentum $ \Pi^{i} $, $ i=1,2,3 $. In temporal gauge, $ \partial_{i}  \Pi^{i}=0$. S-transformation is realized by the unitary operator $ S $ with 
\begin{equation}\label{1q}
S^{-1} \Pi_{i}S=\frac{B_{i}}{2 \pi}\;\;\;\;\;\;\;\;\;S^{-1} B_{i}S=-2 \pi \Pi_{i}\;.
\end{equation}
For the gauge potential eigenstate $ \vert A\rangle $, the S-dual state
\begin{equation}\label{2s}
S \vert A\rangle  =\int DA'\; \exp  \{\frac{i}{2\pi} \int d^{3}x \;\epsilon^{ijk}A'_{i}\partial_{j} A_{k}\}\vert A'  \rangle 
\end{equation}
is the eigenstate of $ \Pi_{i} $ with the eigenvalue $ \frac{1}{2 \pi} \epsilon^{ijk}\partial_{j} A_{k}$. Wilson and 't Hooft operators for the spacial loop $ C $ are given by 
\begin{equation}\label{thj}
W(C) = \exp \{i \oint_{C}ds\; A_{i} \dot{x}^{i} \} = \exp \{i \iint_{\Sigma_{C}}d\sigma^{i}\; B_{i} \}\;\;\;\;\;\;\;\;\;
T(C) = \exp \{ 2 \pi i\iint_{\Sigma_{C}}d\sigma^{i}\; \Pi_{i} \}
\end{equation}
with
\begin{equation}\label{2q}
S^{-1} T(C)S=W(C) \;\;\;\;\;\;\;\;\;S^{-1} W(C) S=T^{+}(C)\;.
\end{equation}
The corresponding magnetic and electric flux operators are 
\begin{equation}
w(C)= \oint_{C} A_{i} (x) dx^{i}   = \iint_{\Sigma_{C}} B_{i} (x) d\sigma^{i}\;\;\;\;\;\;\;\;\;
t(C) = \iint_{\Sigma_{C}} \Pi_{i} (x) d\sigma^{i}  \;,
\end{equation}
satisfying the canonical commutation relation 
\begin{equation}
[w (C), t (C')]= i l(C,C' )\;,
\end{equation}
where $ l(C,C' )  $ is the linking number of the spacial loops $ C $ and $ C' $. 
\begin{equation}
 W(C)=e^{i w(C)}\;\;\;\;\;\;\;\;\;T(C)=e^{2 \pi i t(C)}\;.
\end{equation}
Under the action of $ S $,
\begin{equation}\label{2qa}
S^{-1} t(C)S=\frac{w(C)}{2 \pi} \;\;\;\;\;\;\;\;\;S^{-1} w(C) S=-2 \pi t(C)\;.
\end{equation}
Equations (\ref{1q})-(\ref{2qa}) apply for $ U(1) $ gauge theory with the arbitrary Hamiltonian. If the Hamiltonian is invariant under the action of $ S $ except for a replacement of $ g^{2} $ by $ 1/g^{2} $, the theory will be S-duality invariant.

In Abelian theories, (\ref{1q}) and (\ref{2qa}) are two equivalent basic S-duality transformation rules that could make $ S $ and then (\ref{2s}) entirely determined. Equation (\ref{1q}) does not hold in non-Abelian situation because S-duality transformation can only be definitely defined for gauge invariant operators but $ B_{i} $ and $ \Pi_{i} $ are not gauge invariant any more. On the other hand, loop operators are always gauge invariant, so (\ref{2q}) and (\ref{2qa}) have the non-Abelian extension. Instead of (\ref{2qa}), in this paper, we will only study the non-Abelian version of (\ref{2q}) which could be taken as a semibasic S-duality transformation rule. In Abelian theories, (\ref{2q}) is possible because of (\ref{thj}) as well as the Gauss constraint $ \partial_{i}  \Pi^{i}=0  $ indicating the existence and the exact form of the dual potential. In non-Abelian theories, Wilson and 't Hooft operators are not so symmetric at first sight, so even the validity of (\ref{2q}) needs a proof.

The rest of the paper is organized as follows. Section \ref{sum} is a summary of the main results. Section \ref{The electric} is a review for the electric and the magnetic weight lattices of the gauge group. In Sec. \ref{4}, we give a definition for the gauge invariant 't Hooft operator in canonical formalism and compute the generic commutation relations for Wilson and 't Hooft operators in the arbitrary representations. In Sec. \ref{t-transformation}, we consider the T-transformation of loop operators. In Sec. \ref{S-transformation}, we study the S-transformation of loop operators. The discussion is in Sec. \ref{discussion}.

\section{Summary of main results}\label{sum}

't Hooft operators were first introduced in \cite{Hoo} as the magnetic duals of Wilson operators to detect the phases of gauge theories. As a disorder operator, the 't Hooft operator is usually defined in path integral formalism by specifying a singularity along the loop for fields to be integrated over \cite{10, path}. An explicit realization of the 't Hooft operator in canonical formalism was also given in \cite{onhoo}. In Sec. \ref{4}, based on the 't Hooft operator $ T_{R} (C)$ constructed in \cite{onhoo} labeled by the representation $ R $ of the gauge group, we give a refined 't Hooft operator $\mathcal{T}_{R} (C)  $ in (\ref{par}) which is gauge invariant and moreover, takes a similar form as the Wilson operator $\mathcal{W}_{R} (C)  $, if the latter is written as (\ref{thew}) according to \cite{Wi, 0801, 08011}. We also get the generic canonical commutation relation (\ref{4.26}) for $  \mathcal{W}_{R} (C_{1}) $ and $  \mathcal{T}_{R'} (C_{2}) $ labeled by the arbitrary representations $ R $ and $ R' $. Wilson and 't Hooft operators in representation and the dual representation commute, and especially, when the gauge group is $ U(N) $ which is self-dual, all loop operators commute.

With $ \mathcal{T}_{R}(C)  $ given, we study its T-transformation in Sec. \ref{t-transformation}. It turns out that the 't Hooft operator is indeed multiplied by a Wilson operator as is proposed in \cite{10,tw10}. The obtained Wilson-'t Hooft operator $ [\mathcal{TW}]_{R}(C)   $ is (\ref{twq1}) other than the product $\mathcal{T}_{R}(C)\mathcal{W}_{R}(C)   $. Loop operators  with the arbitrary electric-magnetic weights take the unified form.

In this paper, we mainly focus on theories with the gauge group $ U(N) $, for which Wilson and 't Hooft operators commute. To demonstrate the symmetry between $\mathcal{W}_{R}(C)   $ and $ \mathcal{T}_{R}(C)  $, in Sec. \ref{S-transformation}, we construct their common eigenstate $\vert D  \rangle_{(A',A)}   $ in (\ref{D1}) satisfying (\ref{map11}). $\mathcal{W}_{R}(C)   $ and $ \mathcal{T}_{R}(C)  $ share the same spectrum, so it is indeed possible to construct a unitary operator $ S $ relating the two like (\ref{2q}). We give a tentative construction of $ S $ in (\ref{ho}), keeping in mind that the mapping of loop operators is not enough to uniquely determine $ S $, because, as is already emphasized, it is the flux operators that compose the fundamental gauge invariant observables. All discussions on YM theory are also extended to $ \mathcal{N}=4 $ SYM theory with the gauge group $ U(N) $, where the Wilson and the 't Hooft operators are given by $ \mathcal{W}_{R} (\tau; \lambda^{I}, \lambda^{a}, C)  $ and $ \mathcal{T}_{R} (\tau; \lambda^{I}, \lambda^{a}, C)    $ with $ \tau $ the coupling constant, $ \lambda^{I} $ and $\lambda^{a}  $ the arbitrary periodic scalar and spinor functions on loop $ C $ characterizing the scalar and the fermionic couplings. We show that it is also possible to construct $ S $ making the two kinds of loop operators transformed into each other as in (\ref{opl}). This is consistent with the expectation that S-transformation should make the 't Hooft operator labeled by $ R $ in theory with the coupling constant $ \tau $ mapped into the Wilson operator labeled by $ R $ in theory with the coupling constant $ -1/\tau $ \cite{10}.

Having established the S-duality transformation rule at the kinematical level, we consider the action of $ S $ at the dynamical level. For $ \mathcal{N}=4 $ SYM theory, if the supercharges transform as (\ref{90}) and (\ref{909}), the theory will be duality invariant \cite{SHO}. In Sec. \ref{S-transformation}, we calculate the supersymmetry transformation of the loop operators with $ \lambda^{a} =0$ and show that at least in this case, the constructed $ S $ is consistent with (\ref{90}) and (\ref{909}). As a byproduct, successive action of the supercharges also gives the 't Hooft supermultiplet (\ref{1q21}), which is difficult to construct in path integral formalism by analyzing the singular configurations. In this paper, the 't Hooft operator is defined in canonical formalism, but it is usually studied in path integral formalism \cite{10, path}. For completeness, in Sec. \ref{S-transformation}, we investigate the relation between the two kinds of definitions and show that it is possible to extract $ \mathcal{T}_{R}(C) $ and $ \mathcal{T}_{R}(\tau;\lambda^{I},\lambda^{a},C) $ in canonical formulation from the path integral. Conversely, with the canonical definition of the 't Hooft supermultiplet (\ref{1q21}) given, their path integral definition can also be reconstructed.

In our discussion, we only consider the S-duality transformation of bare operators. The bare operators in the original theory should be mapped into the bare operators in the dual theory. The renormalization is Hamiltonian dependent. For example, starting from (\ref{thj}), the renormalized loop operators in Maxwell theory and Born-Infeld theory are different. It is expected that when combined with the transformation of the coupling constant, the same $ S $ will also make the renormalized operators mapped into each other. S-duality transformation can be implemented along the RG flow, which is also reflected in AdS/CFT, where the type IIB S-duality will make $ \tau $ transformed to $-1/\tau $ along the radial direction of $ AdS_{5} $.

\section{The electric and magnetic weight lattices}\label{The electric}

This section is a brief review of the electric and magnetic weight lattices with the notations and equations that will appear in later discussions listed.

For a semisimple and simply connected group $ G $ with the rank $ r $, $ \{t_{M} | \;  M=1,2,\cdots,\dim G\} $ are generators for the Lie algebra of $ G $ in fundamental representation, among which $\{ H_{A}|\;A=1,2,\cdots,r\}$ are generators of the Cartan subalgebra. $ tr(t_{M}t_{N})=\frac{1}{2}\delta_{MN} $. Simple roots, simple coroots, and fundamental roots are $ r $-dimensional vectors $\{ \vec{\alpha}_{A} |\;A=1,2,\cdots,r\}$, $\{ \vec{\alpha}^{*}_{A} |\;A=1,2,\cdots,r\}$, and $ \{ \vec{\lambda}_{A}|\;A=1,2,\cdots,r \} $,
\begin{equation}\label{1q1h}
2 \vec{\alpha}^{*}_{A}\cdot \vec{\lambda}_{B}= \delta_{AB}\;.
\end{equation}
$\{\vec{\alpha}^{*}_{A}\}  $ and $\{\vec{\lambda}_{A}\}  $ generate the magnetic and the electric weight lattices.

An irreducible representation $ R $ of the group $ G $ is labeled by 
\begin{equation}\label{12}
H_{\vec{m}}=\sum_{A=1}^{r}m_{A}  \vec{\lambda}_{A}\cdot \vec{H}\;,\;\;\;\;\;\;\;\;\;m_{A}\in \mathbb{Z}\;,\;\;\;\;\;m_{A} \geq 0\;, 
\end{equation}
where $ \vec{H}=(H_{1},H_{2},\cdots,H_{r})  $. $  \exp \{4 \pi i H_{\vec{m}}\}=Z$. $ Z $ is a center element of the group. The dual representation $ R^{*} $, which is also an irreducible representation of the GNO dual group $ G^{*} $ \cite{GNO}, is labeled by 
 \begin{equation}
H^{*}_{\vec{m}}=\sum_{A=1}^{r}m_{A}  \vec{\alpha}^{*}_{A}\cdot \vec{H}\;,
\end{equation} 
$  \exp \{4 \pi i H^{*}_{\vec{m}}\}=I  $. According to (\ref{1q1h}),
\begin{equation}\label{1qq1}
tr(H_{\vec{m}}H^{*}_{\vec{m}'})=\frac{1}{4}    \sum_{A=1}^{r} m_{A}m'_{A}\;.
\end{equation}
Moreover, 
\begin{equation}\label{1qq1asf}
tr(H_{\vec{m}}H_{\vec{m}'})=\frac{1}{2}\sum_{A=1}^{r} \sum_{B=1}^{r}  m_{A}m'_{B}\vec{\lambda}_{A}\cdot \vec{\lambda}_{B}\;.
\end{equation}

When $ G=SU(N) $, $ r=N-1 $, $\{ H_{A}|\;A=1,2,\cdots,N-1\}$ can be selected as 
 \begin{equation}
H_{A}=\frac{1}{\sqrt{2A(A+1)}}diag(\underbrace{1,\cdots,1}_{A},-A,\underbrace{0,\cdots,0}_{N-A-1})\;.
\end{equation}
$ |\vec{\alpha}_{A}|^{2}=1 $, $\vec{\alpha}^{*}_{A}=\vec{\alpha}_{A}  $, $  |\vec{\lambda}_{A}|^{2}= \frac{N-1}{2N}$, $ \vec{\lambda}_{A}\cdot \vec{\lambda}_{B} =-\frac{1}{2N} $ for $ A\neq B $. The magnetic weight lattice is a sublattice of the electric weight lattice. For the arbitrary irreducible representation $ R $ and $ R' $ labeled by $ \vec{m} $ and $ \vec{m}' $, from (\ref{1qq1asf}), 
\begin{equation}\label{hm}
tr(H_{\vec{m}}H_{\vec{m}'})=\frac{1}{4}\sum_{A=1}^{r}   m_{A}m'_{A}-\frac{1}{4N}\sum_{A=1}^{r} \sum_{B=1}^{r}  m_{A}m'_{B}\;.
\end{equation}
When $ R $ is the fundamental representation, up to a Weyl transformation, $ \vec{m} $ can be taken as $ \vec{m} =(1,\underbrace{0,\cdots,0}_{N-2})$ with the related 
\begin{equation}\label{aq1}
H  =diag  (\frac{1}{2}-\frac{1}{2N},\underbrace{-\frac{1}{2N},\cdots,-\frac{1}{2N}}_{N-1}) \;,
\end{equation}
$ \exp \{4 \pi i H\}=e^{-2 \pi i/N} I$. The dual representation $ R^{*} $ is associated with
\begin{equation}
H^{*}  =diag(\frac{1}{2},-\frac{1}{2},\underbrace{0,\cdots,0}_{N-2})\;.
\end{equation}

When $G= U(N) $ which is not semisimple but could be locally decomposed as $ SU(N) \times U(1) $, $ r=N $, 
 \begin{equation}
H_{A}=diag(\underbrace{0,\cdots,0}_{A-1},\frac{1}{\sqrt{2}},\underbrace{0,\cdots,0}_{N-A})\;,
\end{equation}
$ A=1,2,\cdots,N $. The fundamental weights and the simple coroots are the same:
 \begin{equation}
\vec{\lambda}_{A}=\vec{\alpha}^{*}_{A}=(\underbrace{0,\cdots,0}_{A-1},\frac{1}{\sqrt{2}},\underbrace{0,\cdots,0}_{N-A})\;.
\end{equation} 
The electric and the magnetic weight lattices are identical. For an irreducible representation $ R $ with $ \vec{m} =(m_{1},\cdots,m_{N})$, 
\begin{equation}\label{24}
H_{\vec{m}}  =H^{*}_{\vec{m}}  =diag(\frac{m_{1}}{2},\cdots,\frac{m_{N}}{2}) \;.
\end{equation}
$ \exp \{4 \pi i H_{\vec{m}}\}= \exp \{4 \pi i H^{*}_{\vec{m}}\}=I $. When $ R $ is the fundamental representation and $ \vec{m} =(1,0,\cdots,0) $,
\begin{equation}\label{aq2}
H =H^{*}  =diag(\frac{1}{2},\underbrace{0,\cdots,0}_{N-1}) \;.
\end{equation}

\section{'t Hooft operator in canonical formalism}\label{4}

In this section, based on the ``bare" 't Hooft operator constructed in \cite{onhoo}, we give a gauge invariant version of the 't Hooft operator in canonical formalism and compute the commutation relations for 't Hooft and Wilson operators in the arbitrary representations of the gauge group.

In canonical quantization formalism, $ 4d $ Yang-Mills (YM) theory with the gauge group $ G$ has the canonical coordinate $ A_{i} $ and the conjugate momentum $ \Pi_{i} $, $ i=1,2,3 $, $A_{i} =A^{M}_{i}  t^{M} $, $\Pi_{i} =\Pi^{M}_{i}  t^{M}  $, $ M=1,2,\cdots, \dim G$. The complete orthogonal basis of the Hilbert space $ \mathcal{H} $ can be selected as the gauge potential eigenstates $ \{\vert A_{i} \rangle | \;\forall \; A\} $. Suppose $ \mathcal{G} $ is the group composed by the local gauge transformation operators $ U $, $ \forall \; U \in   \mathcal{G} $, 
\begin{equation}\label{31}
U  \vert A_{i} \rangle =  \vert U^{-1}A_{i} \rangle =  \vert uA_{i}u^{-1}+iu\partial_{i}u^{-1} \rangle\;,
\end{equation}
where $ u $ is an element of $ G $ in fundamental representation. The physical Hilbert space $  \mathcal{H}_{ph}$ consists of states invariant under the action of $ U $. $\forall \; \vert \psi \rangle \in \mathcal{H}_{ph}  $, $  \forall \; U \in \mathcal{G}  $, $ U\vert \psi \rangle = \vert \psi \rangle $. $ \forall \;  \vert A_{i} \rangle$, the corresponding physical state $ \vert A_{i} \rangle_{ph} \in \mathcal{H}_{ph}$ can be constructed as
\begin{equation}
\vert A_{i} \rangle_{ph}=\int DU \; U  \vert A_{i} \rangle\;.
\end{equation}

For the arbitrary spacial loop $ C $, the Wilson loop in representation $ R $ is given by  
\begin{equation}\label{wii}
\mathcal{W}_{R}(A;C)=\dfrac{1}{d_{R}} tr P\exp \{i \oint_{C}ds\; A_{R}^{i} \dot{x}_{i} \}\;,
\end{equation}
where $ d_{R} $ is the dimension of the representation and $ A^{i}_{R} $ is the gauge potential  in representation $ R $. For the fundamental representation, we will use $A^{i}  $ and $ \mathcal{W} $ with the subscript $ R $ omitted.

Instead of (\ref{wii}), Wilson loop in representation $ R $ labeled by $  \vec{m}$ has an equivalent definition as a path integral over all gauge transformations periodic along the loop $ C $ \cite{Wi, 0801, 08011}:
\begin{equation}\label{ba}
\mathcal{W}_{R}(A;C)=\frac{1}{d_{R}                                                                                                                                                                                                                                                                                                                                                                                                                                                                                                                                                                                                                                                                                                                                                                                                                                                                                                                                                                                                                                                                                                                                                                                                           n_{(R,C)}}\int D_{R}u(s)   \;\exp \{i \oint_{C}ds\; 2tr[H_{\vec{m}} (u^{-1}A_{i}u+iu^{-1}\partial_{i}u) ]\dot{x}^{i} \}\;.
\end{equation}
In (\ref{ba}), all fields are in the fundamental representation with the information on $ R $ encoded in $H_{\vec{m}}  $. The path integral measure $D_{R}u(s) $ is the Haar measure of the group $ G $ times an $ R $-dependent constant. Suppose $ G_{H} $ is the Cartan subgroup of $ G $, then the integrand of (\ref{ba}) is invariant under the action of $ G_{H} $. The constant $ n_{(R,C)} $ in the denominator comes from the integration over the stationary group $ G_{H}  $ along $ C $. So effectively, the path integral is carried over the coset space $ G/G_{H} $ at $ C $. In Appendix \ref{AAA}, a direct proof of (\ref{ba}) is given by performing the path integral explicitly for $ G=SU(2) $.

The action of the Wilson operator $ \mathcal{W}_{R}(C) $ on $ \vert A_{i} \rangle $ is
\begin{equation}
\mathcal{W}_{R}(C)  \vert A_{i} \rangle = \mathcal{W}_{R}(A;C)  \vert A_{i} \rangle\;.
\end{equation}
Based on (\ref{ba}), another operator $ W_{R}(C) $ can be introduced with 
\begin{equation}
W_{R}(C)  \vert A_{i} \rangle = W_{R}(A;C)  \vert A_{i} \rangle\;,
\end{equation}
where 
\begin{equation}
W_{R}(A;C)=\exp \{i \oint_{C}ds\;2 tr(H_{\vec{m}}A_{i} )\dot{x}^{i} \}\;.
\end{equation}
$ \mathcal{W}_{R}(C)   $ can be written in terms of $W_{R}(C)   $:
\begin{equation}\label{thew}
\mathcal{W}_{R}(C)=\frac{1}{d_{R} N_{(R,C)}}\int D_{R}U \;UW_{R}(C)U^{-1}\;. 
\end{equation}
The integration is taken over all of the local gauge transformations $ U \in \mathcal{G} $ in the entire space, but $ W_{R}(C) $ is only affected by $ G/G_{H} $ at the loop $ C $. For $ U $ with $u \in G_{H}$ at $ C $, $ UW_{R}(C)U^{-1}=W_{R}(C) $. The integration over the stationary group of $W_{R}(C)$ gives the divided factor $N_{(R,C)}  $ in the denominator.

To see the equivalence between (\ref{ba}) and (\ref{thew}), consider the situation for $ G=SU(N) $. The gauge transformation periodic on $ C $ gives a closed curve in $ SU(N) $, $ \Pi_{1}(SU(N))= 0$, so the curve can be continuously deformed to $ I $ and the gauge transformation on $ C $ can be covered by $  U \in \mathcal{G}  $. When $G=U(N)  $, $ \Pi_{1}(U(N))=  \mathbb{Z}$; however, the $ U(1) $ part of the transformation periodic on $ C $ will make $W_{R}(C) \rightarrow W_{R}(C)\exp \{4k\pi i \;tr(H_{\vec{m}})\}  =W_{R}(C)\exp \{2km\pi i\}=W_{R}(C) $ thus could be neglected. Equation (\ref{thew}) is still equivalent to (\ref{ba}).

In \cite{Hoo}, the 't Hooft operator $ \mathcal{T} (C)$ is introduced satisfying the canonical commutation relation
\begin{equation}\label{com}
\mathcal{T}(C_{1}) \mathcal{W}(C_{2})  = Z^{-l(C_{1},C_{2})  }   \mathcal{W}(C_{2})  \mathcal{T}(C_{1})  
\end{equation}
with the Wilson operator $ \mathcal{W}(C) $. Here, $\mathcal{T}  $ and $\mathcal{W}  $ stand for $\mathcal{T}_{R}  $ and $\mathcal{W}_{R}  $ with $ R $ the fundamental representation. When $ G=SU(N) $, the center element $ Z=\exp \{2 \pi i/N\} $.

In \cite{onhoo}, an operator $ T_{R}(C) $ satisfying (\ref{com}) was explicitly constructed. The action of $ T_{R}(C) $ on $ \vert A_{i} \rangle$ is given by 
\begin{equation}\label{aqw111}
T_{R}(C)\vert A_{i}\rangle  =\vert T^{-1}_{R}(C)A_{i}\rangle  =\vert \Omega_{\vec{m}}(\Sigma_{C})A_{i}\Omega^{-1}_{\vec{m}}(\Sigma_{C})- H_{\vec{m}}a_{i}(C)\rangle\;. 
\end{equation}
For $ R $ labeled by $H_{\vec{m}}  $, $ \Omega_{\vec{m}}(\Sigma_{C}, x) =  e^{-iH_{\vec{m}}\omega (\Sigma_{C},x)}  $. $ \Sigma_{C} $ is an arbitrary surface with the boundary $ C $, $ \partial \Sigma_{C} =C $. $ \omega (\Sigma_{C},x) $ is the solid angle subtended by the loop $ C $ seen from the point $ x $, and is continuous everywhere except at the surface $\Sigma_{C}  $. 
\begin{equation}\label{sih}
\omega (\Sigma_{C},x)=4\pi\iint_{\Sigma_{C}}d^{2}\sigma_{i}\partial^{x}_{i}D(x-\bar{x}(\sigma))\;, 
\end{equation}
where $D(x)  $ is the Green’s function of the three-dimensional Laplacian, $ -\partial^{2} D(x)=\delta^{3}(x)$.
\begin{equation}\label{101}
a_{i}(C,x)=4\pi\oint_{C} d\bar{x}_{k} \; \epsilon_{kij}\partial^{\bar{x}}_{j}D(x-\bar{x})\;.
\end{equation}
When $ x $ crosses $ \Sigma_{C} $, $ \omega (\Sigma_{C},x) \rightarrow \omega (\Sigma_{C},x)\pm 4\pi$ and accordingly, $ \Omega_{\vec{m}}(\Sigma_{C}, x) \rightarrow  Z^{\pm 1}\Omega_{\vec{m}}(\Sigma_{C}, x)$, since $ e^{-4\pi i H_{\vec{m}}} =Z$. Depending on the different $ \Sigma_{C} $ selected, $  \Omega_{\vec{m}}(\Sigma_{C}, x) $ is determined up to the multiplication of $ Z $, but $  \Omega_{\vec{m}}(\Sigma_{C})A_{i}\Omega^{-1}_{\vec{m}}(\Sigma_{C})$ is only $ C $ dependent.

Away from $ \Sigma_{C} $, $  \Omega_{\vec{m}}(\Sigma_{C}, x)$ is smooth with $i \Omega_{\vec{m}}^{-1}(\Sigma_{C})\partial_{i}\Omega_{\vec{m}}(\Sigma_{C})= H_{\vec{m}}a_{i}(C)$. Locally, $ T_{R}(C)A_{i}$ and $ A_{i} $ are related by a gauge transformation. $  \mathcal{W}(T_{R}(C_{1})A_{i};C_{2})=Z^{-l(C_{1},C_{2})  } \mathcal{W}(A_{i};C_{2}) $. $ T_{R}(C) $ is an operator satisfying \cite{onhoo}
\begin{equation}\label{t}
T_{R}(C_{1}) \mathcal{W}(C_{2})  =Z^{-l(C_{1},C_{2})  } \mathcal{W}(C_{2})  T_{R}(C_{1})  \;.
\end{equation}
When $ G=SU(N) $ and $ R $ is the fundamental representation, $ Z=\exp \{ 2\pi i/N\} $, 
\begin{equation}
T(C_{1}) \mathcal{W}(C_{2})  =\exp \{-\frac{2 \pi i l(C_{1},C_{2})}{N}\}  \mathcal{W}(C_{2})  T(C_{1})  \;.
\end{equation}

When $ e^{-4\pi i H_{\vec{m}}} =I $, which is the situation for $ G=U(N) $ and $ R $ the arbitrary representation or $ G=SU(N) $ and $ R $ a dual representation, $ \Omega_{\vec{m}}(\Sigma_{C}, x) =\Omega_{\vec{m}}(C, x)  $ is a $ C$ dependent continuous function with $   i \Omega_{\vec{m}}^{-1}(C) \partial_{i}\Omega_{\vec{m}}(C) =H_{\vec{m}} a_{i} (C)$ everywhere. Even though, due to the singularity at $ C $, $ H_{\vec{m}} a_{i} (C) $ is not a pure gauge while $ T_{R}(C) $ is not a gauge transformation. For example, when $G= U(1) $, $ H_{\vec{m}}=1/2 $, 
\begin{equation}\label{tc}
T(C)\vert A_{i}\rangle   =\vert A_{i}-\frac{1}{2} a_{i}(C)\rangle  \;.
\end{equation}
The magnetic field $ B^{i} = \epsilon^{ijk}\partial_{j}A_{k} $ transforms as $ B^{i}\rightarrow B^{i}-\frac{1}{2}b^{i}(C) $, where 
\begin{equation}\label{16c}
b^{i}(C,x)=\epsilon^{ijk}\partial_{j}a_{k}(C,x)=4 \pi \oint_{C} d\bar{x}^{i} \; \delta^{3}(x-\bar{x})
\end{equation}
is nonvanishing only at $ C $. $ T(C_{1}) \mathcal{W}(C_{2})  =\mathcal{W}(C_{2})  T(C_{1}) $. $ A_{i} $ and $A_{i}- \frac{1}{2}a_{i}(C) $ have the same Wilson loop but are not the gauge equivalent configurations. $ T(C) $ in (\ref{tc}) is the same as the one defined in (\ref{thj}) when acting on physical states.

When $ G $ is non-Abelian, $T_{R}(C)  $ is not a physical operator. For $ \vert  \psi  \rangle \in \mathcal{H}_{ph} $, $ T_{R}(C) \vert  \psi  \rangle $ may not be a state in $ \mathcal{H}_{ph} $. Consider the action of $ T_{R}(C) $ on a gauge transformation operator $ U $ defined in (\ref{31}), the gauge transformation matrix for $  T^{-1}_{R}(C)UT_{R}(C) $ is $ \Omega_{\vec{m}}(\Sigma_{C})u\Omega^{-1}_{\vec{m}}(\Sigma_{C}) $. $ T^{-1}_{R}(C)UT_{R}(C) $ will be a gauge transformation if $ \Omega_{\vec{m}}(\Sigma_{C})u\Omega^{-1}_{\vec{m}}(\Sigma_{C})  $ is still a single-valued continuous function taking values in $ G$ and approaching $ I $ at infinity, which requires $ [u, H_{\vec{m}}]=0 $ at $ C $, since $\Omega_{\vec{m}}(\Sigma_{C}) $ is singular on $ C $. In this case, $\forall \;  \vert  \psi  \rangle \in \mathcal{H}_{ph} $
\begin{equation}
UT_{R}(C) \vert  \psi  \rangle =T_{R}(C) T_{R}^{-1}(C) UT_{R}(C) \vert  \psi  \rangle  =T_{R}(C)  U' \vert  \psi  \rangle=T_{R}(C)  \vert  \psi  \rangle\;.
\end{equation}
Otherwise, $ UT_{R}(C) \vert  \psi  \rangle  \neq T_{R}(C) \vert  \psi  \rangle $. $ T_{R}(C) \vert  \psi  \rangle $ is a state that would only be affected by the gauge transformation $ G/G_{H} $ at $ C $ thus shares the same stationary group as $  W_{R}(C) $. In this respect, $ T_{R}(C) $ is quite similar with $ W_{R}(C) $.

In parallel with (\ref{thew}), the gauge invariant 't Hooft operator can be constructed as 
\begin{equation}\label{par}
\mathcal{T}_{R}(C) =\frac{1}{d_{R}N_{(R,C)}} \int D_{R}U \;UT_{R}(C)U^{-1}\;.
\end{equation}
From (\ref{t}), 
\begin{equation}
\mathcal{T}_{R}(C_{1}) \mathcal{W}(C_{2})  =Z^{-l(C_{1},C_{2})  }  \mathcal{W}(C_{2})  \mathcal{T}_{R}(C_{1})  \;.
\end{equation}
We get the explicit form of the 't Hooft operator in (\ref{com}).

Equation (\ref{com}) was originally proposed as the commutation relation for loop operators in fundamental representation of $ SU(N) $. It is straightforward to compute the commutation relation for loop operators in arbitrary representations of the group $ G $. Consider $ T_{R}(C_{1}) $ and $ W_{R'}(C_{2}) $ labeled by $ H_{\vec{m}} $ and $ H_{\vec{m}'} $,
\begin{eqnarray}
\nonumber T_{R}(C_{1}) W_{R'}(C_{2})   T^{-1}_{R}(C_{1}) &=& \exp \{i \oint_{C_{2}}ds\;2 tr[H_{\vec{m}'}(A_{i}+H_{\vec{m}}a_{i}(C_{1}) )]\dot{x}^{i} \} \\ &=& W_{R'}(C_{2})  \exp \{8\pi i l(C_{1},C_{2})  tr[H_{\vec{m}'}H_{\vec{m}} ]\}\;,
\end{eqnarray}
\begin{equation}\label{thew111}
T_{R}(C_{1}) W_{R'}(C_{2})=W_{R'}(C_{2})   T_{R}(C_{1}) \exp \{8\pi i l(C_{1},C_{2}) tr[H_{\vec{m}'}H_{\vec{m}} ] \}\;,
\end{equation}
where we have used 
\begin{equation}
\oint_{C_{2}}ds\;a_{i}(C_{1}) \dot{x}^{i}=4\pi l(C_{1},C_{2})\;.
\end{equation}
$ W_{R^{'}}(C_{2}) $ is only affected by the gauge transformation at $ C_{2} $, so instead of (\ref{thew}), the Wilson operator $ \mathcal{W}_{R'}(C_{2}) $ can also be written as 
\begin{equation}\label{thew1}
\mathcal{W}_{R'}(C_{2})=\frac{1}{d_{R'}n_{(R',C_{2})}} \int D_{R'}U(C_{2}) \;[U(C_{2})W_{R^{'}}(C_{2})U^{-1}(C_{2})]\;,
\end{equation}
where $U(C_{2})  $ is the gauge transformation equal to $ I $ away from a thin torus surrounding $ C_{2} $. From (\ref{thew1}) and (\ref{thew111}), 
\begin{eqnarray}\label{329}
\nonumber T_{R}(C_{1})\mathcal{W}_{R'}(C_{2}) &=&\frac{1}{d_{R'}n_{(R',C_{2})}} \int D_{R'}U(C_{2}) \;[T_{R}(C_{1})U(C_{2})T^{-1}_{R}(C_{1})T_{R}(C_{1})W_{R^{'}}(C_{2})U^{-1}(C_{2}) ]\\ \nonumber &=& \frac{1}{d_{R'}n_{(R',C_{2})}} \int D_{R'}U(C_{2}) \;[T_{R}(C_{1})U(C_{2})T^{-1}_{R}(C_{1})W_{R^{'}}(C_{2})  T_{R}(C_{1})U^{-1}(C_{2})T^{-1}_{R}(C_{1})\\ \nonumber &&T_{R}(C_{1})\exp \{8\pi i l(C_{1},C_{2}) tr[H_{\vec{m}'}H_{\vec{m}} ] \}]\\ &=&  \mathcal{W}_{R'}(C_{2})T_{R}(C_{1})\exp \{8\pi i l(C_{1},C_{2}) tr[H_{\vec{m}'}H_{\vec{m}} ] \}\;,
\end{eqnarray}
where we have used the fact that $ T_{R}(C_{1})U(C_{2})T^{-1}_{R}(C_{1}) $ is still a gauge transformation equal to $ I $ away from the torus surrounding $ C_{2} $. Let 
\begin{equation}
\exp \{ i L(R,R';C_{1},C_{2})  \} := \exp \{8\pi i l(C_{1},C_{2}) tr[H_{\vec{m}'}H_{\vec{m}} ] \} \;,
\end{equation}
from (\ref{par}) and (\ref{329}), 
\begin{equation}\label{4.26}
\mathcal{T}_{R}(C_{1})\mathcal{W}_{R'}(C_{2})=\mathcal{W}_{R'}(C_{2})\mathcal{T}_{R}(C_{1})\exp \{ i L(R,R';C_{1},C_{2})  \}\;.
\end{equation}
This is the generic commutation relation for loop operators in arbitrary representations.

When $ R $ and $ R' $ are irreducible representations of the group $ G $ and the dual group $ G^{*} $, from (\ref{1qq1}), $ \exp \{ i L(R,R';C_{1},C_{2})  \} =1  $,
\begin{equation}\label{54}
\mathcal{T}_{R}(C_{1})\mathcal{W}_{R'}(C_{2})=\mathcal{W}_{R'}(C_{2})\mathcal{T}_{R}(C_{1})\;.
\end{equation}
When $ G=SU(N) $, according to (\ref{hm}),
\begin{equation}\label{515}
 \mathcal{T}_{R}(C_{1})\mathcal{W}_{R'}(C_{2})= \mathcal{W}_{R'}(C_{2})\mathcal{T}_{R}(C_{1})\exp \{-\frac{2\pi i l(C_{1},C_{2}) }{N}\sum_{A=1}^{r} \sum_{B=1}^{r}  m_{A}m'_{B}\}\;.
\end{equation}
Especially, when $ R $ and $ R' $ are both the fundamental representations labeled by $ H $,  
\begin{equation}
\mathcal{T}(C_{1})\mathcal{W}(C_{2})=\mathcal{W}(C_{2})\mathcal{T}(C_{1})\exp \{- \frac{2 \pi  i l(C_{1},C_{2})}{N} \}\;.
\end{equation}
When $ G=U(N) $, for the arbitrary $ R $ and $ R' $, from (\ref{24}), $  \exp \{ i L(R,R';C_{1},C_{2})  \} =1 $,
\begin{equation}
\mathcal{T}_{R}(C_{1})\mathcal{W}_{R'}(C_{2})=\mathcal{W}_{R'}(C_{2})\mathcal{T}_{R}(C_{1})\;.
\end{equation}

\section{T-transformation of loop operators}\label{t-transformation}

S-duality transformation is generated by T-transformation and S-transformation. In canonical quantization formalism, T-transformation could be realized by a unitary operator $ g(A)  =\exp \{-\frac{iX(A)}{2 \pi}\} $ \cite{T} with 
\begin{equation}
X(A) =\frac{1}{2} \int d^{3}x\; \epsilon^{ijk}tr(A_{i}\partial_{j}A_{k}-\frac{2i}{3}A_{i}A_{j}A_{k})
\end{equation}
the Chern-Simons functional. 
\begin{equation}
g(A)\Pi_{i} g^{-1}(A) =  \Pi_{i}+\frac{B_{i}}{2\pi}\;\;\;\;\;\;\;\;\;g(A)A_{i} g^{-1}(A) =  A_{i}\;,
\end{equation}
where $ B^{i} =\frac{1}{2} \epsilon^{ijk}  F_{jk}$.

The Wilson operator is invariant under the action of $ g $, $ g(A)\mathcal{W}_{R}(C)  g^{-1}(A) =\mathcal{W}_{R}(C)  $. As for the T-transformation of the 't Hooft operator, we can compute the action of $ T_{R}(C) $ on $ X(A)  $:
\begin{equation}
T_{R}(C)X(A)T^{-1}_{R}(C) = X(T_{R}(C)A)\;,
\end{equation}
where $T_{R}(C)A_{i} = \Omega_{\vec{m}}^{-1}(\Sigma_{C})A_{i}\Omega_{\vec{m}}(\Sigma_{C})+H_{\vec{m}}a_{i}(C) $. Direct calculation gives 
\begin{equation}
X(T_{R}(C)A)= X(A)+4\pi \oint_{C}ds\; tr(H_{\vec{m}}A_{i} )\dot{x}^{i}\;,
\end{equation}
so 
\begin{equation}
T_{R}(C)g(A)T^{-1}_{R}(C) = W^{-1}_{R}(C)g(A)\;,
\end{equation}
or equivalently,
\begin{equation}
g(A)T_{R}(C) g^{-1}(A)= T_{R}(C)W_{R}(C)\;. 
\end{equation}
Under the T-transformation, $ T_{R}(C) $ is multiplied by $ W_{R}(C) $ in the same representation. On the other hand, T-transformation of the gauge invariant 't Hooft operator $ \mathcal{T}_{R}(C) $ is 
\begin{equation}\label{twq1}
g(A)\mathcal{T}_{R}(C) g^{-1}(A)= \frac{1}{d_{R} N_{(R,C)}}\int D_{R}U \;U[T_{R}(C)W_{R}(C)]U^{-1} :=[\mathcal{TW}]_{R}(C) \;,
\end{equation}
where $ [\mathcal{TW}]_{R}(C) $ could be taken as the Wilson-'t Hooft operator originally proposed in path integral formulation \cite{10}. Here, $ W $ and $ T $ in $[\mathcal{TW}]_{R}  $ are both labeled by the representation $ R $. $ [\mathcal{TW}]_{R}(C)   $ is different from the double trace operator $ \mathcal{T}_{R}(C)\mathcal{W}_{R}(C)  $ unless the gauge group is $ U(1) $ so that $\mathcal{W}(C) =W(C)  $, $ \mathcal{T}(C) =T(C)  $.

\section{S-transformation of loop operators}\label{S-transformation}

When the gauge group is $ U(N) $, S-transformation is expected to make the 't Hooft operator $ \mathcal{T}_{R}(C) $ in theory with the coupling constant $\tau $ mapped into the Wilson operator $  \mathcal{W}_{R}(C) $ in theory with the coupling constant $ -1/\tau $ \cite{10,path}. For it to be possible, two kinds of operators should be equivalent. In this section, we will study the spectrum and eigenstates of the 't Hooft operator in YM theory as well as the $ \mathcal{N} =4$ SYM theory. We will show that it is possible to construct a unitary operator $ S $ relating the loop operators as in (\ref{unn}) and (\ref{opl}).

So S-transformation could be realized at the kinematical level. At the dynamical level, if $ S $ could also make the Hamiltonian with the coupling constant $ \tau $ transformed into the Hamiltonian with the coupling constant $-1/ \tau $, the theory will be S-duality invariant. For $ \mathcal{N} =4$ SYM theory, the condition for the S-duality invariance is that supercharges should transform as (\ref{90}) and (\ref{909}) \cite{SHO}. We will calculate the supersymmetry variations of the loop operators and provide the evidence for the $ U(1)_{Y} $ transformation of the supercharges under the action of $ S $.

\subsection{An exercise in one-dimensional quantum mechanics}

Before getting into the technical details, we will exemplify the strategy in one-dimensional quantum mechanics. In one-dimensional quantum mechanics, the position and momentum operators are $ X $ and $ P $, 
\begin{equation}
[X,P]=i \;,\;\;\;\;\;\;\;\;\;\;e^{ i X}  e^{2 \pi i P} =  e^{2 \pi i P} e^{ i X} .
\end{equation}

It is possible to construct the common eigenstates of $e^{ i X}   $ and $e^{2 \pi i P}   $. As the eigenstates of $ e^{ i X}   $, $ \{ \vert x \rangle|\;\forall \;x \in \mathbb{R} \} $ can be divided as
\begin{equation}
 \{ \vert x \rangle| \;\forall \; x\in \mathbb{R}\}  = \cup_{\hat{x} \in [-\pi, \pi)} E(\hat{x})
\end{equation}
with
\begin{equation}
E(\hat{x}) :=     \{ \vert \hat{x}+2 k\pi \rangle| \;\forall \; k \in \mathbb{Z}\}
\end{equation}
generating $ \mathcal{H}[E(\hat{x})] $, the eigenspace of $ e^{ i X}     $ with the eigenvalue $ e^{ i \hat{x}} $. Common eigenstate of $e^{ i X}   $ and $e^{2 \pi i P}   $ can be constructed in each $ \mathcal{H}[E(\hat{x})] $:
\begin{equation}
 \vert D \rangle_{(\hat{x},\hat{x}')} =\sum_{k} \exp \{\frac{i}{2\pi}(\frac{\hat{x}'}{2}+2k \pi)^{2}\}  \vert \hat{x}+2k \pi \rangle\;.
\end{equation}
\begin{equation}
e^{ i X} \vert D \rangle_{(\hat{x},\hat{x}')} =e^{ i \hat{x}} \vert D \rangle_{(\hat{x},\hat{x}')} \;,\;\;\;\;\;\;\;\;\; e^{2\pi i P}  \vert D \rangle_{(\hat{x},\hat{x}')}  = e^{ i \hat{x}'}  \vert D \rangle_{(\hat{x},\hat{x}')} \;,\;\;\;\;\;\;\;\;\; \hat{x},\hat{x}' \in [-\pi, \pi)\;.
\end{equation}
Aside from $\{ \vert x \rangle|\;\forall \;x \in \mathbb{R} \}   $ and $ \{ \vert p \rangle|\;\forall \;p \in \mathbb{R} \}  $, we get another set of complete orthogonal basis $\{ \vert D \rangle_{(\hat{x},\hat{x}')} |\;\forall \;\hat{x},\hat{x}' \in [-\pi, \pi) \}     $ for $ \mathcal{H} $ parametrized by points in a square. $ \{\hat{x}' \;|\; \hat{x}' \in [-\pi, \pi)\} \sim \{k \;|\;k \in  \mathbb{Z}  \} $, so $ \{(\hat{x},\hat{x}')\;|\;\hat{x},\hat{x}' \in [-\pi, \pi) \} \sim  \{(\hat{x},k) \;|\;\hat{x} \in [-\pi, \pi), k \in  \mathbb{Z} \} \sim \{\hat{x}+2 k\pi  \;|\; \hat{x}+2k \pi  \in \mathbb{R}\} $. The three sets of bases have the same degrees of freedom.

The duality operator $ U $ in one-dimensional quantum mechanics should make 
\begin{equation}\label{ukf}
U^{-1}e^{2 \pi  iP} U=e^{ i X}\;,\;\;\;\;\;\;\;\;\;\;U^{-1}e^{ i X}U= e^{ -2 \pi iP} 
\end{equation}
and then 
\begin{equation}\label{degg}
U \vert D \rangle_{(\hat{x},\hat{x}')} =e^{i \theta (\hat{x},\hat{x}')} \vert D \rangle_{(-\hat{x}',\hat{x})}\;.
\end{equation}
As for the action of $ U $ on $ \{ \vert x \rangle| \;\forall \; x\in \mathbb{R}\}    $, generically, we will have  
\begin{equation}\label{deg}
 U \vert x \rangle=\vert \mathcal{D} \rangle_{x} =\int_{-\pi}^{\pi} d\hat{x}' \;e^{ i h(\hat{x}',x)}\vert D \rangle_{(\hat{x}',x)}\;.
\end{equation}

In (\ref{degg}) and (\ref{deg}), $e^{i \theta (\hat{x},\hat{x}')} $ and $  e^{ i h(\hat{x}',x)}$ are undetermined. Equation (\ref{ukf}) is not enough to fix $ U $ due to the degeneracies in the spectra of $  e^{ i X} $ and $ e^{2 \pi  i P}  $.   A stronger requirement is
\begin{equation}
U^{-1}P U=\frac{X}{2 \pi}  \;,\;\;\;\;\;\;\;\;\;\;U^{-1}XU= -2 \pi P\;,
\end{equation}
which will enforce $ \vert  \mathcal{D} \rangle_{x} $ to be the momentum eigenstate,
\begin{equation}
\vert  \mathcal{D} \rangle_{x} = \int^{\infty}_{-\infty}dx' \; e^{\frac{ i x x'}{2 \pi}}   \vert x' \rangle\;. 
\end{equation}
$e^{ i h(\hat{x}',x)}  $ and $  e^{i \theta (\hat{x}',\hat{x})}  $ are then fixed as
\begin{equation}
e^{ i h(\hat{x}',x)}= e^{-\frac{i x (x-4\hat{x}')}{8 \pi} } \;,\;\;\;\;\;\;\;\;\;\;
 e^{i \theta (\hat{x}',\hat{x})} =e^{\frac{i}{8\pi}(\hat{x}^{2}-4\hat{x} \hat{x}'-\hat{x}'^{2})}    \;.
\end{equation}

\subsection{Spectrum and eigenstates of the 't Hooft operator in YM theory}

Consider YM theory with the gauge group $ U(N) $, for the arbitrary irreducible representations $ R_{1} $, $  R_{2}$ and the arbitrary spacial loops $ C_{1} $, $ C_{2}$, 
\begin{equation}
\mathcal{T}_{R_{2}}(C_{2})\mathcal{W}_{R_{1}}(C_{1})=\mathcal{W}_{R_{1}}(C_{1})\mathcal{T}_{R_{2}}(C_{2})\;.
\end{equation}
We may construct the common eigenstates of $ \mathcal{T}_{R_{2}}(C_{2})$ and $\mathcal{W}_{R_{1}}(C_{1}) $.

Eigenstates of $\mathcal{W}_{R}(C)  $ are $ \{\vert A \rangle|\; \forall \; A\}    $, composing the complete orthogonal basis for the Hilbert space $ \mathcal{H} $. 
\begin{equation}
\mathcal{W}_{R}(C)\vert A \rangle = \mathcal{W}_{R}(A;C)\vert A \rangle\;.
\end{equation}
$\forall \; U  \in \mathcal{G}  $, $ \forall \;C' $, $U\mathcal{W}_{R}(C) =\mathcal{W}_{R}(C)U $, $ T(C')\mathcal{W}_{R}(C) =\mathcal{W}_{R}(C)T(C') $. $ T(C) $ and $ U $ generate a loop group $ \mathcal{L} $ commuting with $ \mathcal{W}_{R}(C) $.
\begin{equation}
\mathcal{L} := \{  U_{n}T( C_{n-1})U_{n-1}\cdots T( C_{2})U_{2}T( C_{1})U_{1} |\; \forall\;U_{i} \in \mathcal{G}, \forall\;C_{i},\forall\; n\} \;.
\end{equation}
Each loop has the assumed orientation. With the orientation reversed, $T(C)\rightarrow T^{-1}( C)   $. When $ C $ shrinks to a point, $T( C)   $ approaches the identity $ I $, so elements in $ \mathcal{L} $ can all be continuously deformed to $ I $. $ \mathcal{L}  $ is an infinite dimensional simply continuous group. $ \forall\; L \in  \mathcal{L}$, $ L\mathcal{W}_{R}(C) =\mathcal{W}_{R}(C)L $, $  \mathcal{W}_{R}(LA;C)=\mathcal{W}_{R}(A;C)$.

The action of $  \mathcal{L}   $ could make $ \{\vert A \rangle|\; \forall \; A\}  $ divided into the equivalent classes. $\vert A \rangle  $ and $\vert A' \rangle  $ belong to the same class if $\exists\;  L \in  \mathcal{L} $, $\vert A' \rangle  =L \vert A \rangle  $.
\begin{equation}\label{32}
\{\vert A \rangle|\; \forall \;A\} =\cup_{\hat{A}}E(\hat{A})\;,
\end{equation}
where
 \begin{equation}
E(\hat{A}):=\{L\vert   \hat{A}\rangle|\; \forall\; L \in \mathcal{L}\}
 \end{equation}
is an equivalent class with $ \vert  \hat{A}\rangle $ the arbitrary element in it. $ E(\hat{A}) $ generates $ \mathcal{H}[E(\hat{A}) ]   $ which is the eigenspace of $\mathcal{W}_{R} (C)     $ with the eigenvalue $\mathcal{W}_{R} (\hat{A};C)  $.

For the arbitrary representation $ R $, $ T_{R}(C) $ can always be decomposed as the product of $ T(C) $ and $ U $, so $T_{R}(C)  \in  \mathcal{L}  $. $ \forall \; U \in \mathcal{G}$, $ \forall\; \vert   A\rangle \in E(\hat{A})$, $ U T_{R}(C) U^{-1} \vert   A\rangle \in E(\hat{A})$ and then, $   \mathcal{T}_{R} (C)  \vert   A\rangle \in  \mathcal{H}[E(\hat{A}) ]$. $ \mathcal{H} [E(\hat{A}) ]  $ is invariant under the action of $  \mathcal{T}_{R} (C)   $.

$ \mathcal{H} $ can be decomposed as 
\begin{equation}
 \mathcal{H} = \oplus_{\hat{A}} \; \mathcal{H} [E(\hat{A}) ] \;. 
\end{equation}
Common eigenstates of $  \mathcal{T}_{R} (C)  $ and $   \mathcal{W}_{R'} (C') $ can be constructed in each $ \mathcal{H}[E(\hat{A}) ] $. Let
\begin{equation}\label{D1}
\vert D  \rangle_{(A',A)} = \int dL\; g^{-1}(L A)\vert L A'  \rangle\;,
\end{equation}
$\vert D  \rangle_{(A',A)} \in   \mathcal{H}[E(A') ]  $. The integration is taken over the group $ \mathcal{L} $ with $ dL $ the invariant measure, $ d(LL')=d(L'L)=dL $. 
\begin{equation}\label{D1j}
\forall \; L \in \mathcal{L}\;,\;\;\;\;\;\;\;\;\;\;\;\;\;\vert D  \rangle_{(A',A)} =\vert D  \rangle_{(LA',LA)} \;.
\end{equation}
$ g(A)=\exp \{-\frac{iX(A)}{2\pi}\} $, where $ X(A) $ is the Chern-Simons functional. 
\begin{equation}
g(T_{R}(C)A)=W^{-1}_{R}(A;C)g(A)\;. 
\end{equation}
$ g(UA)=g(A) $, so
\begin{equation}
U\vert D  \rangle_{(A',A)} = \int dL\; g^{-1}(L A)\vert U^{-1} L A'  \rangle = \int d(U^{-1}L)\; g^{-1}( U^{-1} L A)\vert U^{-1} L A'  \rangle=\vert D  \rangle_{(A',A)}\;,
\end{equation}
$ \vert D  \rangle_{(A',A)}  \in \mathcal{H}_{ph} $.

The action of $ \mathcal{T}_{R} (C)$ on $\vert D  \rangle_{(A',A)}  $ is given by
\begin{eqnarray}
\nonumber \mathcal{T}_{R} (C)\vert D  \rangle_{(A',A)} &=& \frac{1}{d_{R} N_{(R,C)}}\int D_{R}U \; UT_{R}(C)\vert D  \rangle_{(A',A)} \\ &=&\nonumber
\frac{1}{d_{R} N_{(R,C)}}\int D_{R}U \;U \int dL\; g^{-1}(LA) \vert T^{-1}_{R}(C)LA'  \rangle\\ &=&\nonumber
\frac{1}{d_{R} N_{(R,C)}}\int D_{R}U \;U \int d(T^{-1}_{R}(C)L)\; g^{-1}(T_{R}(C)  T^{-1}_{R}(C) LA) \vert T^{-1}_{R}(C)LA'  \rangle\\ &=&\nonumber
\frac{1}{d_{R} N_{(R,C)}}\int D_{R}U \;U \int dL\; g^{-1}(T_{R}(C)   LA) \vert LA'  \rangle\\ &=&\nonumber
\frac{1}{d_{R} N_{(R,C)}}\int D_{R}U \; \int dL\; W_{R}(  LA;C)g^{-1}(  LA) \vert U^{-1}LA'  \rangle\\ &=&
\nonumber\frac{1}{d_{R} N_{(R,C)}}\int D_{R}U \; \int dL\; W_{R}(U  LA;C)g^{-1}(  LA) \vert LA'  \rangle\\ &=&
\nonumber\int dL\; \mathcal{W}_{R}(LA;C)g^{-1}(  LA) \vert LA'  \rangle\\ &=&\mathcal{W}_{R}(A;C)
\int dL\;  g^{-1}(  LA) \vert LA'  \rangle  =\mathcal{W}_{R}(A;C)\vert D  \rangle_{(A',A)}\;.
\end{eqnarray} 
So $ \forall \; R $, $ \forall \; C $, $ \vert D  \rangle_{(A',A)} $ is the common eigenstate of $ \mathcal{W}_{R} (C)   $ and $ \mathcal{T}_{R} (C)   $ with
\begin{equation}\label{map11}
\mathcal{W}_{R} (C)   \vert D  \rangle_{(A',A)} =\mathcal{W}_{R}(A';C)\vert D  \rangle_{(A',A)} \;,\;\;\;\;\;\;\;\;\;\;\mathcal{T}_{R} (C)   \vert D  \rangle_{(A',A)} =\mathcal{W}_{R}(A;C)\vert D  \rangle_{(A',A)} \;.
\end{equation}
$ \vert D  \rangle_{(A',A)}  $ is $ R $ and $ C $ independent by construction.

The conjugation of $ \mathcal{T}_{R} (C) $ is 
\begin{equation}
\mathcal{T}_{R}^{+} (C)=\frac{1}{d_{R} N_{(R,C)}}\int D_{R}U\; UT_{R}^{-1}(C)U^{-1}
\end{equation}
with
\begin{eqnarray}
\nonumber \mathcal{T}_{R}^{+} (C)\vert D  \rangle_{(A',A)}&=& \int dL\; \frac{1}{d_{R} N_{(R,C)}}\int D_{R}U\;W_{R}^{-1}(UL A;C)g^{-1}(L A) \vert L A'  \rangle \\ \nonumber &=& \int dL\; \mathcal{W}_{R}^{*} (L A;C)g^{-1}(L A) \vert L A'  \rangle \\ &=&\mathcal{W}_{R}^{*} ( A;C) \vert D  \rangle_{(A',A)}\;,
\end{eqnarray}
where $\mathcal{W}_{R}^{*} (A;C)= \frac{1}{d_{R} N_{(R,C)}}\int D_{R}U\;W_{R}^{-1}(UA;C)$ is the complex conjugate of $  \mathcal{W}_{R} (A;C)$.

For $ |D \rangle_{(A_{1},A_{2})}  $ and $  |D \rangle_{(A'_{1},A'_{2})} $,
\begin{equation}
_{(A_{1},A_{2})}\langle D |\mathcal{T}_{R} (C)|D \rangle_{(A'_{1},A'_{2})}=\mathcal{W}_{R}(A_{2};C) _{(A_{1},A_{2})}\langle D | D \rangle_{(A'_{1},A'_{2})}=\mathcal{W}_{R}(A'_{2};C) _{(A_{1},A_{2})}\langle D | D \rangle_{(A'_{1},A'_{2})}\;,
\end{equation}
so $ _{(A_{1},A_{2})}\langle D | D \rangle_{(A'_{1},A'_{2})}= 0$ if $\mathcal{W}_{R}(A_{1};C)  \neq \mathcal{W}_{R}(A'_{1};C)  $ or $\mathcal{W}_{R}(A_{2};C)  \neq \mathcal{W}_{R}(A'_{2};C)  $.

In Appendix \ref{BBB}, it is shown that $ \{  \vert D  \rangle_{(A',A)}|\;\forall \;A    \} $ composes the complete basis for $ \mathcal{H}_{ph}[E(A') ] $, so altogether $ \{  \vert D  \rangle_{(A',A)}|\;\forall \;A',A    \}  $ composes the over-complete basis for $  \mathcal{H}_{ph}$. By over-completeness, we have taken into account of the possible degeneracies such as (\ref{D1j}). Especially, when $G= U(1) $, $ \vert D  \rangle_{(A',LA)} = \vert D  \rangle_{(A',A)} $ so $ \{  \vert D  \rangle_{(A',A)}|\;\forall \;A',A    \} =   \{  \vert D  \rangle_{(\hat{A}',\hat{A})}|\;\forall \;\hat{A}',\hat{A}    \}$ with $ \hat{A}',  \hat{A} $ parametrizing the equivalent classes.

When the gauge group is $ U(1) $, $ \mathcal{T} (C)=T(C) $, $ \mathcal{W} (C)=W(C)  $, 
\begin{equation}
g(A)=\exp \{-\frac{i}{2 \pi}\int d^{3}x \; \frac{1}{2}\epsilon^{ijk}A_{i} \partial_{j}A_{k}\}=\exp \{-\frac{i}{2 \pi}\int d^{3}x \; \frac{1}{2}A_{i} B^{i}\}\;.
\end{equation}
For $ \vert A\rangle=\vert L\hat{A}\rangle $ and $L=  U_{n}T( C_{n-1})\cdots U_{2} T( C_{1})U_{1}  $, $B_{i}=\hat{B}_{i}+\frac{1}{2} \sum^{n-1}_{k=1} b_{i} ( C_{k})  $ with $ b_{i} $ given by (\ref{16c}).
\begin{eqnarray}\label{la}
\nonumber g(L \hat{A}) &=& \prod^{n-1}_{k=1}W^{-1}(\hat{A}; C_{k})g(\hat{A})=g^{-1}(\hat{A}) \exp \{\frac{i}{2\pi }\int d^{3}x \; \hat{A}^{i} (-\frac{1}{2}\sum^{n-1}_{k=1} b_{i}(  C_{k})-\hat{B}_{i})\} \\  &=& g^{-1}(\hat{A}) \exp \{-\frac{i}{2\pi}\int d^{3}x \; \hat{A}^{i} B_{i}\} =g^{-1}(\hat{A}) \exp \{-\frac{i}{2\pi}\int d^{3}x \; A^{i} \hat{B}_{i}\}\;,
\end{eqnarray}
where we have used 
\begin{equation}
W(A; C) =\exp \{ i \oint_{ C} ds\; A_{i}\dot{x}^{i} \}= \exp \{\frac{i}{4 \pi}\int d^{3}x \; A_{i} b^{i}( C)\}\;.
\end{equation}
As a result,
\begin{equation}\label{u1u}
\vert D  \rangle_{(\hat{A}',\hat{A})} =g(\hat{A})\exp  \{\frac{i}{2\pi} \int d^{3}x \;(\hat{A}_{i}-\hat{A}'_{i})\hat{B}^{i} \} \sum_{\vert A' \rangle \in E(\hat{A}')} \exp  \{\frac{i}{2\pi} \int d^{3}x \;A'_{i}\hat{B}^{i} \}\vert A'  \rangle\;,
\end{equation}
which is just (\ref{2s}) with the integration over $ A' $ restricted in $ \vert A' \rangle \in E(\hat{A}') $ in order to be the common eigenstate of $T(C)  $ and $ W(C) $.

\subsection{Mapping of the loop operators}

Equation (\ref{map11}) exhibits a symmetry between $\mathcal{T}_{R} (C)$ and $\mathcal{W}_{R} (C)$. One may construct a unitary operator $ S $ with
\begin{equation}\label{unn}
S^{-1}\mathcal{T}_{R}(C) S=\mathcal{W}_{R}(C)\;,\;\;\;\;\;\;\;\;\;S^{-1}\mathcal{W}_{R}(C)S= \mathcal{T}_{R}^{+}(C)\;,\;\;\;\;\;\;\;\forall \; R\; , \;\forall \; C\;.
\end{equation}
$ S^{2} $ is the charge conjugation operator with  
\begin{equation}
S^{-2}\mathcal{T}_{R}(C) S^{2}=\mathcal{T}_{R}^{+}(C)\;,\;\;\;\;\;\;\;\;\;S^{-2}\mathcal{W}_{R}(C)S^{2}= \mathcal{W}_{R}^{+}(C)\;,
\end{equation}
\begin{equation}
 S^{2} \vert A \rangle=\vert A^{C} \rangle  \;.
\end{equation}
For $A =A^{M}  t^{M} $, $ M=1,2,\cdots, \dim G$, the charge conjugate configuration is $A^{C} =-A^{M}  t^{M*}   $ \cite{1982, 19823}.
\begin{equation}
\mathcal{W}_{R}(A^{C};C) = \mathcal{W}_{R}^{*}(A;C)\;,\;\;\;\;\;\;\;\;\; g(A^{C}) =g(A)\;.
\end{equation}
The charge conjugation of the gauge potential $ LA $ is $ L^{*}A^{C}$. If $  L=U_{n}T( C_{n-1})\cdots T(  C_{1})U_{1}  $, $ L^{*}=U^{*}_{n}T^{-1}( C_{n-1})\cdots T^{-1}(  C_{1})U^{*}_{1}$. For $ \vert D  \rangle_{(A',A)} =\int dL\; g^{-1}(LA)\vert LA'\rangle $,
\begin{eqnarray}
\nonumber S^{2}\vert D  \rangle_{(A',A)}  &=& \int dL\; g^{-1}(LA)S^{2}\vert LA'\rangle=\int dL\; g^{-1}(LA)\vert L^{*}A'^{C}\rangle \\  &=& \int dL^{*}\; g^{-1}(L^{*}A^{C})\vert L^{*}A'^{C}\rangle =\int dL\; g^{-1}(LA^{C})\vert LA'^{C}\rangle=\vert D  \rangle_{(A'^{C},A^{C})}\;.
\end{eqnarray}

The operator $ S $ can be defined through its action on the basis $ \{  \vert D  \rangle_{(A',A)}|\;\forall \;A',A    \}   $. From (\ref{unn}), $ S\vert D  \rangle_{(A',A)} $ should satisfy
\begin{equation}\label{mn}
\mathcal{T}_{R}(C) S\vert D  \rangle_{(A',A)}=\mathcal{W}_{R}(A';C)S\vert D  \rangle_{(A',A)}\;,\;\;\;\;\;\;\;\;\;\mathcal{W}_{R}(C)S\vert D  \rangle_{(A',A)}= \mathcal{W}_{R}(A^{C};C)S\vert D  \rangle_{(A',A)}\;,
\end{equation}
which, however, is not enough to fix it. A trial solution is 
\begin{equation}\label{ho}
 S\vert D  \rangle_{(A',A)}  =\exp \{\frac{i}{4\pi}\int d^{3}x \; tr[(B_{i}-B'_{i})(A^{i}-A'^{i})]\}
\vert D  \rangle^{*}_{(A^{C},A'^{C})}
\end{equation}
with
\begin{equation}
\vert D  \rangle^{*}_{(A^{C},A'^{C})} =\int dL\; g(LA'^{C})\vert LA^{C}\rangle
\end{equation}
composing another set of over-complete bases for $ \mathcal{H}_{ph}  $. $ \vert D  \rangle^{*}_{(A^{C},A'^{C})} = \vert D  \rangle^{*}_{((LA)^{C},(LA')^{C})} $, which is consistent with (\ref{D1j}). When $ G=U(1) $, $\vert D  \rangle_{(A',A)}   $ and $ S $ are given by (\ref{u1u}) and (\ref{2s}), in which case, (\ref{ho}) holds exactly.

As for the action of $ S $ on $ \{\vert A \rangle_{ph}|\; \forall \; A\} $, generically, we will have 
\begin{equation}\label{hhh}
S\vert A \rangle_{ph}=\int DA'\;h(A,A') \vert D  \rangle_{(A',A)}:=\vert \mathcal{D} \rangle_{A}
\end{equation}
for some $ h(A,A') $ to be determined. 
\begin{equation}
\mathcal{T}_{R} (C)  \vert \mathcal{D} \rangle_{A} =\mathcal{W}_{R} (A;C) \vert \mathcal{D} \rangle_{A}\;. 
\end{equation}
$\{\vert A \rangle_{ph}|\; \forall \; A\}  $ and $ \{\vert \mathcal{D} \rangle_{A}|\; \forall \; A \} $ compose two sets of complete orthogonal bases for $ \mathcal{H}_{ph} $. 
\begin{equation}
S \vert A \rangle_{ph}=\vert \mathcal{D} \rangle_{A}\;,\;\;\;\;\;\;\;\;\; S\vert \mathcal{D} \rangle_{A}= \vert A^{C} \rangle_{ph}\;.
\end{equation}

The spectrum of $ \mathcal{T}_{R}(C) $ and $  \mathcal{W}_{R}(C)$ are highly degenerate, so (\ref{unn}) can only make $ S $ determined up to $ S \sim VS $ with $ [V,\mathcal{T}_{R}(C) ]=[V,\mathcal{W}_{R}(C)] =0$, which is also reflected in the ambiguity of $ h$ in (\ref{hhh}). On the other hand, flux operators have the reduced degeneracy. Let $ \mathcal{T}_{R}(C)=e^{2 \pi i t_{R}(C)} $, $\mathcal{W}_{R}(C)=e^{i w_{R}(C)}  $ with $t_{R}(C)  $ and $ w_{R}(C) $ the corresponding electric and the magnetic flux operators, the mapping 
\begin{equation}\label{un}
S^{-1}t_{R}(C) S=\frac{w_{R}(C)}{2 \pi}\;,\;\;\;\;\;\;\;\;\;S^{-1}w_{R}(C)S=-2\pi t_{R}^{+}(C)
\end{equation}
may fix $ S $ completely just as the $ U(1) $ case.

\subsection{Modified 't Hooft operator}

The standard 't Hooft operator $ \mathcal{T}_{R}(C) $ satisfies the commutation relation 
\begin{equation}
\mathcal{T}_{R}(C_{1})\mathcal{W}_{R'}(C_{2}) = \mathcal{W}_{R'}(C_{2})\mathcal{T}_{R}(C_{1})\exp \{ i L(R,R';C_{1},C_{2})  \}\;.
\end{equation}
For an operator $ Y(C) $ built from $ A $, if $[T_{R}(C),Y(C)]=0  $, $ Y^{-1}(C)=Y^{+}(C) $, let
\begin{equation}
 \mathcal{T}'_{R}(C)=\frac{1}{d_{R} N_{(R,C)}}\int D_{R}U\;  UT_{R}'(C)U^{-1} =\frac{1}{d_{R} N_{(R,C)}}\int D_{R}U\; U[T_{R}(C) Y(C)]U^{-1}\;,
\end{equation}
then $ \mathcal{T}'_{R}(C) $ still satisfies
\begin{equation}
\mathcal{T}'_{R}(C_{1})\mathcal{W}_{R'}(C_{2}) = \mathcal{W}_{R'}(C_{2})\mathcal{T}'_{R}(C_{1})\exp \{ i L(R,R';C_{1},C_{2})  \}\;.
\end{equation}

If there is a gauge invariant operator $ K(A) $ with
\begin{equation}\label{tf}
T_{R}^{-1} (C)K(A)T_{R}(C)= K(T_{R}(C)A) =Y(A,C)K(A)\;,
\end{equation}
then
\begin{equation}\label{sta}
K(A)T_{R}(C)K^{-1}(A) =T_{R}(C)Y(C)\;,
\end{equation}
\begin{equation}\label{sta1}
K(A)\mathcal{T}_{R} (C)K^{-1}(A)=\frac{1}{d_{R} N_{(R,C)}}\int D_{R}U\;  U[T_{R}(C)Y(C)]U^{-1}=\mathcal{T}_{R}' (C) \;. 
\end{equation}
$ \mathcal{T}_{R}' (C) $ and $ \mathcal{T}_{R} (C) $ are similar.

Equation (\ref{tf}) could be taken as an integration equation for $K(A)  $ with the integrable condition
\begin{equation}\label{ing}
Y(T_{R}(C_{2})A,C_{1})Y^{-1}(A,C_{1})=Y(T_{R}(C_{1})A,C_{2})Y^{-1}(A,C_{2})
\end{equation}
coming from the commutation relation $ [T'_{R}(C_{1}),T'_{R}(C_{2}) ]=0$. If (\ref{ing}) is not satisfied, $T'_{R}(C_{1})$ and $T'_{R}(C_{2})  $ do not commute. Even though, $ T'_{R} (C) $ and $ T_{R} (C) $ can still be similar, but are related by a $ C $ dependent operator $ K(A,C)$.

When $ Y(A,C)=W_{R}(A,C) $, $ K(A)=g(A) $, (\ref{sta}) and (\ref{sta1}) become 
\begin{equation}
 g(A)T_{R}(C)g^{-1}(A)=T_{R}(C)W_{R}(C)
\end{equation}
and 
\begin{equation}
 g(A) \mathcal{T}_{R} (C)g^{-1}(A)=\mathcal{T}_{R}' (C)\;,
\end{equation}
where 
\begin{equation}
\mathcal{T}'_{R} (C)=\frac{1}{d_{R} N_{(R,C)}}\int D_{R}U\;  U[T_{R}(C)W_{R}(C)]U^{-1}\;.
\end{equation}
This is the T-transformation of the 't Hooft operator.

\subsection{'t Hooft operator in path integral formalism and canonical formalism}\label{formalism}

The 't Hooft operator is usually defined in path integral formalism \cite{path}. In this subsection, we will study the relation between the canonical and path integral formulations of the 't Hooft operator and extract the former from the latter.

In path integral formalism, the 't Hooft operator is introduced by expanding quantum fields around the singular configurations. For the globally defined gauge potential $ A_{\mu} $, the Bianchi identity is automatically satisfied:
\begin{equation}
\epsilon^{\mu\nu\rho\sigma}D_{\nu}F_{\rho\sigma}=0\;,
\end{equation}
where 
\begin{equation}
F_{\mu\nu}=\partial_{\mu}A_{\nu}-\partial_{\nu}A_{\mu}-i[A_{\mu},A_{\nu}]\;.
\end{equation}
With the current $ j^{\mu} $ given, 
\begin{equation}\label{2}
\epsilon^{\mu\nu\rho\sigma}D_{\nu}F_{\rho\sigma}=j^{\mu}
\end{equation}
also has the solution $ G_{\mu} $, which is not globally defined. With $ G_{\mu} $ plugged in, (\ref{2}) could reduce to 
\begin{equation}\label{3h}
2\epsilon^{\mu\nu\rho\sigma}\partial_{\nu}\partial_{\rho}G_{\sigma} =j^{\mu}\;.
\end{equation}

To get the 't Hooft operator $ \mathcal{T}_{R}(C) $ for a spacial loop $ C $ at the time $ t=t_{0} $, $ j^{\mu} $ is taken to be 
\begin{equation}\label{5577}
 j^{0}=0 \;,\;\;\;\;\;\;\; j^{i} = 2H_{\vec{m}} b^{i}\delta (t-t_{0})\;.
\end{equation}
The corresponding $ G_{\mu} $ is solved as
\begin{equation}\label{557}
 G_{i} =0 \;,\;\;\;\;\;\;\;  G_{0} =-H_{\vec{m}} \omega \delta (t-t_{0})\;.
\end{equation}
$ b^{i}=\epsilon^{ijk}\partial_{j}a_{k} $. $ b^{i}$, $ \omega $ and $ a_{i}  $ are given by (\ref{16c}), (\ref{sih}) and (\ref{101}), respectively. $ \omega (\Sigma_{C},x) $ jumps $ 4 \pi $ when $ x $ crosses $ \Sigma_{C} $, but $e^{i \omega}  $ is continuous except for the singularity at $ C $. Formally, $a_{i}=-i e^{-i \omega}\partial_{i}  e^{i \omega}$ even if $ a_{i} $ is not a pure gauge, so locally, $ a_{i} =\partial_{i}\omega  $. With (\ref{557}) and (\ref{5577}) plugged in (\ref{3h}), we do have
\begin{eqnarray}
\nonumber && 2\epsilon^{ijk0}\partial_{j}\partial_{k}G_{0}  =2H_{\vec{m}}\epsilon^{ijk}\partial_{j}\partial_{k}\omega \delta (t-t_{0})=2H_{\vec{m}}\epsilon^{ijk}\partial_{j}a_{k}\delta (t-t_{0})=2H_{\vec{m}} b^{i}\delta (t-t_{0})=j^{i}\\&& 2\epsilon^{0ijk}\partial_{i}\partial_{j}G_{k}  = 0 \;.
\end{eqnarray}

In (\ref{3h}), $ G_{\mu} $ is determined up to the addition of an arbitrary globally defined gauge potential $  \tilde{A}_{\mu}$. The generic solution of (\ref{2}) is $ A_{\mu} =G_{\mu}+\tilde{A}_{\mu}$. \begin{equation}
F_{\mu\nu}=\tilde{F}_{\mu\nu} + \tilde{D}_{\mu}G_{\nu}-\tilde{D}_{\nu}G_{\mu}-i[G_{\mu},G_{\nu}]\;,
\end{equation}
where $ \tilde{F}_{\mu\nu} = \partial_{\mu}\tilde{A}_{\nu}-\partial_{\nu}\tilde{A}_{\mu}-i[\tilde{A}_{\mu},\tilde{A}_{\nu}]$, $ \tilde{D}_{\mu}f= \partial_{\mu}f-i[\tilde{A}_{\mu},f]$. If the background $ G_{\mu} $ is taken to be (\ref{557}), 
\begin{equation}
F_{ij}=\tilde{F}_{ij}\;,\;\;\;\;\;\;\;\;\;
F_{i0}=\tilde{F}_{i0}+\tilde{D}_{i}G_{0}\;.
\end{equation}

Consider the YM theory with the gauge group $ G $ and the coupling constant $ \tau $, $ \tau = \tau_{1}+i \tau_{2} $. According to the relation between the YM theory and $ D3 $ branes in type IIB string theory, $\tau_{2}=1/g_{s}  $, where $ g_{s} $ is the type IIB string coupling constant. The gauge coupling is $ g=\sqrt{2 \pi g_{s}} $. $ 1/g^{2} = \frac{\tau_{2}}{2 \pi} $. In the presence of the 't Hooft operator, the gauge potential should be $  A_{\mu} =G_{\mu}+\tilde{A}_{\mu} $ and the action becomes 
\begin{eqnarray}
\nonumber S&=&\frac{\tau_{2}}{2\pi}\int d^{4}x\;tr(-\dfrac{1}{4}F_{\mu\nu}F^{\mu\nu}+\dfrac{\tau_{1}}{8\tau_{2}}\epsilon^{\mu\nu\rho\sigma}F_{\mu\nu}F_{\rho\sigma})\\ \nonumber &=&\frac{\tau_{2}}{2\pi}\int d^{4}x\;   tr(-\dfrac{1}{4}\tilde{F}_{\mu\nu}\tilde{F}^{\mu\nu}+\dfrac{\tau_{1}}{8\tau_{2}}\epsilon^{\mu\nu\rho\sigma}\tilde{F}_{\mu\nu}\tilde{F}_{\rho\sigma}+\tilde{F}_{i0}\tilde{D}^{i}G_{0} +\dfrac{1}{2}\tilde{D}_{i}G_{0}\tilde{D}^{i}G_{0} -\dfrac{\tau_{1}}{2\tau_{2}}\epsilon^{ijk}\tilde{F}_{ij}\tilde{D}_{k}G_{0})\\  &=&\tilde{S}+\frac{1}{2\pi}
\int d^{3}x\;tr[-\tau_{2}  \tilde{F}_{i0}\tilde{D}^{i}(H_{\vec{m}}\omega) + \dfrac{\tau_{2} }{2}\delta(0)\tilde{D}_{i}(H_{\vec{m}}\omega)\tilde{D}^{i}(H_{\vec{m}}\omega)+\tau_{1}\tilde{B}_{i}\tilde{D}^{i}(H_{\vec{m}}\omega)]\;,
\end{eqnarray}
where
\begin{equation}
 \tilde{S}=\frac{\tau_{2}}{2\pi}\int d^{4}x\;   tr(-\dfrac{1}{4}\tilde{F}_{\mu\nu}\tilde{F}^{\mu\nu}+\dfrac{\tau_{1}}{8\tau_{2}}\epsilon^{\mu\nu\rho\sigma}\tilde{F}_{\mu\nu}\tilde{F}_{\rho\sigma})
 \end{equation} 
is the YM action for $\tilde{A}  $. The path integral measure is $D A  = D \tilde{A}$. $\tilde{A}  $ is the dynamical field with the modified action $S  $. In temporal gauge, $  \tilde{\Pi}_{i} =- \frac{\tau_{2}}{2\pi}\tilde{F}_{i0}+\frac{\tau_{1}}{2\pi}\tilde{B}_{i}$, so replacing $ \tilde{S} $ by $ S $ amounts to adding the operator 
\begin{equation}
T_{R}'(C)=\exp \{i \int d^{3}x\;tr[\tilde{\Pi}_{i}\tilde{D}^{i}(H_{\vec{m}}\omega)+\dfrac{\tau_{2}}{4 \pi}\delta(0)\tilde{D}_{i}(H_{\vec{m}}\omega)\tilde{D}^{i}(H_{\vec{m}}\omega)]\}
\end{equation}
into the path integral. $ T_{R}'(C)=T_{R}(C)Y(C) $, where 
\begin{equation}\label{try}
T_{R}(C)=\exp \{i \int d^{3}x\;tr[\tilde{\Pi}_{i}\tilde{D}^{i}(H_{\vec{m}}\omega)]\}=\exp \{i \int d^{3}x\;tr(H_{\vec{m}}\tilde{\Pi}^{i}a_{i}-i[\tilde{\Pi}^{i},\tilde{A}_{i}]H_{\vec{m}}\omega)\}
\end{equation}
is the 't Hooft operator in canonical formalism constructed in \cite{onhoo}, 
\begin{equation}
Y(C) = \exp \{i \int d^{3}x\;tr[\dfrac{\tau_{2}}{4 \pi}\delta(0)\tilde{D}_{i}(H_{\vec{m}}\omega)\tilde{D}^{i}(H_{\vec{m}}\omega)]\}
\end{equation}
is an operator constructed from $ \tilde{A} $. $ T_{R}'(C)$ and $T_{R}(C) $ are equivalent if the suitable $ K(A) $ can be found as is in (\ref{sta}).

$   A_{\mu} =G_{\mu}+\tilde{A}_{\mu}$, under the gauge transformation $ U $, 
\begin{equation}
G_{\mu}\rightarrow uG_{\mu}u^{-1}\;\;\;\;\;\;\;\;\;\tilde{A}_{i}\rightarrow u\tilde{A}_{i}u^{-1}+iu\partial_{i}u^{-1}\;\;\;\;\;\;\;\;\;j_{i}\rightarrow uj_{i}u^{-1}\;.
\end{equation}
To preserve the gauge invariance, the path integral should cover the background $ uG_{\mu}u^{-1}$ for the arbitrary $ U $. The obtained gauge invariant 't Hooft operator is 
\begin{equation}
\mathcal{T}_{R}'(C) =\frac{1}{d_{R}N_{(R,C)}}\int D_{R}U\; UT_{R}'(C)U^{-1}  \;.
\end{equation}

Under the T-transformation, $ \tilde{\Pi}_{i}\rightarrow \tilde{\Pi}_{i}+\frac{\tilde{B}_{i}}{2 \pi} $. From (\ref{try}), 
\begin{eqnarray}
\nonumber T_{R}(C)&\rightarrow & \exp \{i \int d^{3}x\;tr[\tilde{\Pi}_{i}\tilde{D}^{i}(H_{\vec{m}}\omega)]+\frac{1}{2\pi}tr[\tilde{B}_{i}\tilde{D}^{i}(H_{\vec{m}}\omega)]\}\\ \nonumber&=&\exp \{i \int d^{3}x\;tr[\tilde{\Pi}_{i}\tilde{D}^{i}(H_{\vec{m}}\omega)]\}\exp \{i \oint_{C}ds\;2 tr(H_{\vec{m}}\tilde{A}_{i} )\dot{x}^{i} 
\}=T_{R}(C)W_{R}(C)\;,\\
\end{eqnarray} 
while
\begin{equation}
\mathcal{T}_{R}(C) \rightarrow  \frac{1}{d_{R}N_{(R,C)}}\int D_{R}U\; U[T_{R}(C)W_{R}(C)]U^{-1}\;.
\end{equation}
This is the manifestation of the T-transformation rule in path integral formalism.

In canonical formulation, the 't Hooft operator, in parallel with the Wilson operator, is determined by the field content with no dynamical information involved. But in path integral formalism, the 't Hooft operator is action dependent. $\mathcal{T}_{R}'(C)= \frac{1}{d_{R}N_{(R,C)}}\int D_{R}U\; U[T_{R}(C)Y(C)]U^{-1}  $, where $ Y(C) $ depends on the concrete form of the action. It is expected that $ \mathcal{T}_{R}'(C) $ and $ \mathcal{T}_{R}(C) $ are equivalent.

The above discussion can be extended to $ \mathcal{N} =4$ SYM theory. The Lagrangian is 
\begin{eqnarray}\label{ll}
\nonumber L &=& \frac{\tau_{2}}{2\pi}tr \{-\frac{1}{4}F_{\mu\nu}F^{\mu\nu}+\dfrac{\tau_{1}}{8\tau_{2}}\epsilon^{\mu\nu\rho\sigma}F_{\mu\nu}F_{\rho\sigma}-i\bar{\Psi}^{a}\bar{\sigma}^{\mu}D_{\mu}\Psi_{a}-\frac{1}{2}D_{\mu}\Phi^{I}D^{\mu}\Phi^{I}\\&&
+\;\frac{1}{2}C^{ab}_{I} \Psi_{a} [\Phi^{I},\Psi_{b}]+\frac{1}{2} \bar{C}_{Iab} \bar{\Psi}^{a} [\Phi^{I},\bar{\Psi}^{b} ]+\frac{1}{4} [\Phi^{I},\Phi^{J}]^{2}\}\;,
\end{eqnarray}
$\mu,\nu =0,1,2,3  $, $a,b=1,2,3,4  $, $ I,J=1,2,\cdots,6 $. Following \cite{path}, aside from the gauge potential $ A_{\mu} $, background fields for the scalar $ \Phi^{I}  $ and the fermion $ \Psi^{a} $ can also be introduced: 
\begin{equation}
A_{\mu} =G_{\mu}+\tilde{A}_{\mu}\;\;\;\;\;\;\;\; \Phi^{I} =\phi^{I} +\tilde{\Phi}^{I}\;\;\;\;\;\;\;\; \Psi^{a} =\psi^{a}+\tilde{\Psi}^{a}\;.
\end{equation}
$ \phi^{I}  $ and $ \psi^{a} $ satisfy
\begin{equation}
\partial^{\mu}\partial_{\mu}\phi^{I}(x) =2(\frac{\tau_{2}}{2\pi})^{-\frac{1}{2}}H_{\vec{m}}\oint_{C}ds\; \lambda^{I}(s) \delta^{4}[x-\bar{x}(s)]
\end{equation}
and 
\begin{equation}
\sigma^{\mu}\partial_{\mu}\psi^{a}(x) =-2(\frac{\tau_{2}}{2\pi})^{-\frac{1}{2}}H_{\vec{m}}\oint_{C}ds\; \lambda^{a}(s) \delta^{4}[x-\bar{x}(s)]
\end{equation}
for some functions $ \lambda^{I}(s)  $ and $ \lambda^{a}(s) $ on loop $ C $. The Lagrangian can be expanded as
\begin{eqnarray}
\nonumber L&=& \tilde{L}+tr\{\frac{1}{2\pi}  (\tau_{2}\tilde{F}_{i0} -\tau_{1}\tilde{B}_{i})\tilde{D}^{i}G_{0}- \frac{\tau_{2}}{2\pi} (
\bar{\tilde{\Psi}}^{a}\bar{\sigma}^{0}[G_{0},\tilde{\Psi}_{a}]+i\partial_{0}\tilde{\Phi}^{I}[G_{0},\tilde{\Phi}^{I}])
 \\ \nonumber &-& \frac{\tau_{2}}{2\pi}  \partial_{\mu}\phi^{I}\partial^{\mu}\tilde{\Phi}_{I}+ \frac{i\tau_{2}}{2\pi}  \partial_{\mu} \bar{\psi}^{a}\bar{\sigma}^{\mu}\tilde{\Psi}_{a} -\frac{i\tau_{2}}{2\pi}   \bar{\tilde{\Psi}}^{a}\bar{\sigma}^{\mu}\partial_{\mu}\psi_{a}+r(\tau;\tilde{A}_{\mu},\tilde{\Phi}^{I},\tilde{\Psi}^{a};G_{\mu},\phi^{I},\psi^{a})\}\\\nonumber &\sim& \tilde{L}+tr\{\frac{1}{2\pi}(\tau_{2}\tilde{F}_{i0} -\tau_{1}\tilde{B}_{i})\tilde{D}^{i}G_{0}-\frac{\tau_{2}}{2\pi}(
\bar{\tilde{\Psi}}^{a}\bar{\sigma}^{0}[G_{0},\tilde{\Psi}_{a}]+i\partial_{0}\tilde{\Phi}^{I}[G_{0},\tilde{\Phi}^{I}])
  \\ \nonumber &+& 2 (\frac{\tau_{2}}{2\pi})^{\frac{1}{2}}\oint_{C}ds\; H_{\vec{m}}\tilde{\Phi}_{I}\lambda^{I}(s) \delta^{4}[x-\bar{x}(s)]+2i (\frac{\tau_{2}}{2\pi})^{\frac{1}{2}} \oint_{C}ds\;H_{\vec{m}}(\tilde{\Psi}_{a}^{+}-\tilde{\Psi}_{a}) \lambda^{a}(s) \delta^{4}[x-\bar{x}(s)]  \\  &+& r(\tau;\tilde{A}_{\mu},\tilde{\Phi}^{I},\tilde{\Psi}^{a};G_{\mu},\phi^{I},\psi^{a})\}\;.
\end{eqnarray}

Canonical quantization of $ \tilde{L} $ in temporal gauge gives 
\begin{equation}
\tilde{\Pi}_{i}= -\frac{\tau_{2}}{2\pi}\tilde{F}_{i0} +\frac{\tau_{1}}{2\pi}\tilde{B}_{i}\;,\;\;\;\;\;\tilde{\Pi}^{I}=\frac{\tau_{2}}{2\pi}  \partial_{0}\tilde{\Phi}^{I}  \;,\;\;\;\;\;\tilde{\Pi}^{a} =\frac{i\tau_{2}}{2\pi} \bar{\tilde{\Psi}}^{a} \bar{\sigma}^{0}\;.
\end{equation}
So adding the background fields amounts to adding the operator 
\begin{eqnarray}
\nonumber  &&T_{R}'(\tau;\lambda^{I},\lambda^{a},C)\\ \nonumber &=&  \exp \{i \int d^{4}x\; tr(-\tilde{\Pi}_{i}\tilde{D}^{i}G_{0}+i \tilde{\Pi}^{a}   [G_{0},\tilde{\Psi}_{a}]-i\tilde{\Pi}^{I}[G_{0},\tilde{\Phi}^{I}]+2(\frac{\tau_{2}}{2\pi})^{\frac{1}{2}}\oint_{C}ds\; H_{\vec{m}}\tilde{\Phi}_{I}\lambda^{I}(s) \delta^{4}[x-\bar{x}(s)] \\
 \nonumber &&  +\;    2i(\frac{\tau_{2}}{2\pi})^{\frac{1}{2}} \oint_{C}ds\;H_{\vec{m}}(\tilde{\Psi}_{a}^{+}-\tilde{\Psi}_{a}) \lambda^{a}(s) \delta^{4}[x-\bar{x}(s)]+r(\tau;\tilde{A}_{\mu},\tilde{\Phi}^{I},\tilde{\Psi}^{a};G_{\mu},\phi^{I},\psi^{a}))\}  \\ \nonumber &=&  \exp \{i \int d^{3}x\; tr(\tilde{\Pi}_{i}\tilde{D}^{i}(H_{\vec{m}}\omega)-i \tilde{\Pi}^{a}   [H_{\vec{m}}\omega,\tilde{\Psi}_{a}]+i\tilde{\Pi}^{I}[H_{\vec{m}}\omega,\tilde{\Phi}^{I}])\}  \\ \nonumber &&  \exp \{i (\frac{\tau_{2}}{2\pi})^{\frac{1}{2}}\oint_{C}ds\;2 tr(H_{\vec{m}}\tilde{\Phi}_{I})\lambda^{I}  \}\exp \{- (\frac{\tau_{2}}{2\pi})^{\frac{1}{2}}\oint_{C}ds\; 2tr[H_{\vec{m}}(\tilde{\Psi}_{a}^{+}-\tilde{\Psi}_{a}) ]\lambda^{a} \} \\ \nonumber &&      \exp \{i\int d^{4}x\;r(\tau;\tilde{A}_{\mu},\tilde{\Phi}^{I},\tilde{\Psi}^{a};G_{\mu},\phi^{I},\psi^{a})    \} \\  &=&  T_{R}(C)W_{R}^{\tilde{\Phi}}(\tau;\lambda^{I},C)W_{R}^{\tilde{\Psi}}(\tau;\lambda^{a},C)Y(\tau;\lambda^{I},\lambda^{a},C)
\end{eqnarray} 
into the path integral. 
\begin{equation}
T_{R}(C)=\exp \{i \int d^{3}x\; tr(\tilde{\Pi}_{i}\tilde{D}^{i}(H_{\vec{m}}\omega)-i \tilde{\Pi}^{a}   [H_{\vec{m}}\omega,\tilde{\Psi}_{a}]+i\tilde{\Pi}^{I}[H_{\vec{m}}\omega,\tilde{\Phi}^{I}])\} 
\end{equation}
is the 't Hooft operator generating the singular gauge transformation in $ \mathcal{N} =4$ SYM theory.
\begin{equation}
W_{R}^{\tilde{\Phi}}(\tau;\lambda^{I},C)= \exp \{i (\frac{\tau_{2}}{2\pi})^{\frac{1}{2}}\oint_{C}ds\; 2tr(H_{\vec{m}}\tilde{\Phi}_{I})\lambda^{I}  \}
\end{equation}
and 
\begin{equation}
W_{R}^{\tilde{\Psi}}(\tau;\lambda^{a},C)= \exp \{- (\frac{\tau_{2}}{2\pi})^{\frac{1}{2}} \oint_{C}ds\; 2tr[H_{\vec{m}}(\tilde{\Psi}_{a}^{+}-\tilde{\Psi}_{a}) ]\lambda^{a} \}
\end{equation}
are Wilson loops for $ \tilde{\Phi} $ and $ \tilde{\Psi}$.
\begin{equation}
Y(\tau;\lambda^{I},\lambda^{a},C)=\exp \{i \int d^{4}x\;r(\tau;\tilde{A}_{\mu},\tilde{\Phi}^{I},\tilde{\Psi}^{a};G_{\mu},\phi^{I},\psi^{a}) \}
\end{equation}
is an operator constructed from $\tilde{A}_{i},\tilde{\Phi}^{I},\tilde{\Psi}^{a}  $, whose form depends on action as well as the explicit solutions for $ \phi^{I} $ and $  \psi^{a}$.

Under the local gauge transformation $ U $, 
\begin{eqnarray}
\nonumber &&  H_{\vec{m}}\omega \rightarrow u (H_{\vec{m}}\omega) u^{-1}\;\;\;\;\;\;\;\;\;\phi^{I} \rightarrow u\phi^{I}u^{-1}\;\;\;\;\;\;\;\;\;\psi^{a} \rightarrow u\psi^{a} u^{-1}\;   \\&&
\tilde{A}_{i}\rightarrow u\tilde{A}_{i}u^{-1}+iu\partial_{i}u^{-1}\;\;\;\;\;\;\;\;\;\tilde{\Phi}^{I} \rightarrow u\tilde{\Phi}^{I}u^{-1}\;\;\;\;\;\;\;\;\;\tilde{\Psi}^{I} \rightarrow u\tilde{\Psi}^{I}u^{-1} \;.
\end{eqnarray}
The path integral should cover the background $\{u (H_{\vec{m}}\omega) u^{-1} ,u\phi^{I}u^{-1},u\psi^{a} u^{-1}\} $ for all of $ U $ and the final gauge invariant 't Hooft operator is
\begin{equation}
 \mathcal{T}_{R}'(\tau;\lambda^{I},\lambda^{a},C) =\frac{1}{d_{R}N_{(R,C)}}\int D_{R}U\; UT_{R}'(\tau;\lambda^{I},\lambda^{a},C)U^{-1}\;.
 \end{equation}

\subsection{Spectrum and eigenstates of 't Hooft operator in $  \mathcal{N}=4$ SYM theory}

In this subsection, we will construct the canonical 't Hooft operators in $  \mathcal{N}=4$ SYM theory and calculate their spectra and eigenstates.

In canonical quantization formalism, for $ \mathcal{N} =4$ SYM theory with the coupling constant $ \tau =\tau_{1}+i\tau_{2}$ and the gauge group $ G $, the canonical coordinate is $ \Lambda :=(A_{i},\Phi^{I},\Psi^{a}) $ with the conjugate momentum $ (\Pi^{i},\Pi_{I},\Pi_{a}) $. The generic supersymmetric Wilson operator labeled by $ R$ is \cite{35q, 36q, 0510}
\begin{equation}\label{957}
\mathcal{W}_{R}(\tau;\lambda^{I},\lambda^{a},C)=\dfrac{1}{d_{R}} tr P\exp \{i \oint_{C}ds\; (A_{R}^{i} \dot{x}_{i}+ (\frac{\tau_{2}}{2\pi})^{\frac{1}{2}} \Phi_{R}^{I} \lambda_{I}+(\frac{\tau_{2}}{2\pi})^{\frac{1}{2}}  \Psi_{R}^{a}\lambda_{a}) \}\;,
\end{equation}
where $\lambda^{I}(s)  $ and $\lambda^{a}(s)  $ are scalar and spinor functions on loop $ C $, $ A_{R}, \Phi_{R},\Psi_{R}$ are fields in representation $ R$. To preserve some amount of supersymmetry locally or globally, $\lambda^{a}=0 $, $\lambda^{I} $ and $\dot{x}_{i}  $ should satisfy the particular constraints \cite{35q, 36q, 37q}. In (\ref{957}), $\lambda^{I} $ and $\lambda^{a} $ are arbitrary, since after all, non-BPS operators also exist and have the S-dual.

Similar with (\ref{thew}), (\ref{957}) can also be equivalently written as 
\begin{equation}
\mathcal{W}_{R}(\tau;\lambda^{I},\lambda^{a},C)=\frac{1}{d_{R}N_{(R,C)}}\int D_{R}U\; UW_{R}(\tau;\lambda^{I},\lambda^{a},C)U^{-1}\;,
\end{equation}
where
\begin{equation}
W_{R}(\tau;\lambda^{I},\lambda^{a},C)=W_{R}^{A}(C)W_{R}^{\Phi}(\tau;\lambda^{I},C)W_{R}^{\Psi}(\tau;\lambda^{a},C)\;,
\end{equation}
\begin{eqnarray}
\nonumber && W_{R}^{A}(C)=\exp \{i \oint_{C}ds\; 2tr(H_{\vec{m}}A_{i})\dot{x}^{i} \}\;,\\\nonumber && W_{R}^{\Phi}(\tau;\lambda^{I},C)=\exp \{i (\frac{\tau_{2}}{2\pi})^{\frac{1}{2}} \oint_{C}ds\; 2tr(H_{\vec{m}}\Phi_{I} )\lambda^{I}\}\;,
  \\ &&
 W_{R}^{\Psi}(\tau;\lambda^{a},C)=\exp \{i (\frac{\tau_{2}}{2\pi})^{\frac{1}{2}}\oint_{C}ds\; 2tr(H_{\vec{m}}\Psi_{a} )\lambda^{a} \}\;.
\end{eqnarray} 
Suppose $ \vert \Lambda \rangle := \vert A_{i},\Phi^{I},\Psi^{a} \rangle$ is the common eigenstate of $ (A_{i},\Phi^{I},\Psi^{a})  $, 
\begin{equation}
\mathcal{W}_{R}(\tau;\lambda^{I},\lambda^{a},C)\vert \Lambda \rangle=\mathcal{W}_{R}(\Lambda;\tau;\lambda^{I},\lambda^{a},C)\vert \Lambda \rangle\;.
\end{equation}

The corresponding supersymmetric 't Hooft operator is 
\begin{equation}\label{De3}
\mathcal{T}_{R}(\tau;\lambda^{I},\lambda^{a},C)=\frac{1}{d_{R}N_{(R,C)}}\int D_{R}U\; UT_{R}(\tau;\lambda^{I},\lambda^{a},C)U^{-1}
\end{equation}
with
\begin{equation}
T_{R}(\tau;\lambda^{I},\lambda^{a},C)=T_{R}(C)W_{R}^{\Phi}(\tau;\lambda^{I},C)W_{R}^{\Psi}(\tau;\lambda^{a},C)\;.
\end{equation}
As the generalization of (\ref{aqw111}) in $ \mathcal{N} =4$ SYM theory, the action of $ T_{R}(C) $ on $\vert A_{i},\Phi^{I},\Psi^{a} \rangle   $ is
\begin{equation}
T_{R}(C)\vert A_{i},\Phi^{I},\Psi^{a} \rangle = \vert \Omega_{\vec{m}}(\Sigma_{C})A_{i}\Omega^{-1}_{\vec{m}}(\Sigma_{C})-H_{\vec{m}}a_{i}(C),\Omega_{\vec{m}}(\Sigma_{C})\Phi^{I}\Omega^{-1}_{\vec{m}}(\Sigma_{C}),\Omega_{\vec{m}}(\Sigma_{C})\Psi^{a}\Omega^{-1}_{\vec{m}}(\Sigma_{C}) \rangle\;,
\end{equation}   
where $ \Omega_{\vec{m}}(\Sigma_{C})A_{i}\Omega^{-1}_{\vec{m}}(\Sigma_{C}) $, $ \Omega_{\vec{m}}(\Sigma_{C})\Phi^{I}\Omega^{-1}_{\vec{m}}(\Sigma_{C}) $ and $\Omega_{\vec{m}}(\Sigma_{C})\Psi^{a}\Omega^{-1}_{\vec{m}}(\Sigma_{C})   $ are only $ C $ dependent. As we can see in Sec. \ref{formalism}, (\ref{De3}) is the most generic spacial 't Hooft operator that could be constructed in path integral formalism by specifying the singularities for fundamental fields.

Loop operators given above satisfy the commutation relation
\begin{equation}
T_{R}(C_{1}) \mathcal{W}_{R'}(\tau;\lambda_{2}^{I},\lambda_{2}^{a},C_{2})  = \mathcal{W}_{R'}(\tau;\lambda_{2}^{I},\lambda_{2}^{a},C_{2})  T_{R}(C_{1})  \exp \{ i L(R,R';C_{1},C_{2})  \}
\end{equation}
as well as 
\begin{equation}
\mathcal{T}_{R}(\tau;\lambda_{1}^{I},\lambda_{1}^{a},C_{1})  \mathcal{W}_{R'}(\tau;\lambda_{2}^{I},\lambda_{2}^{a},C_{2})  = \mathcal{W}_{R'}(\tau;\lambda_{2}^{I},\lambda_{2}^{a},C_{2})  \mathcal{T}_{R}(\tau;\lambda_{1}^{I},\lambda_{1}^{a},C_{1})\exp \{ i L(R,R';C_{1},C_{2})  \} \;.
\end{equation}
Especially, when $ G=U(N) $, 
\begin{equation}
T_{R}(C_{1}) \mathcal{W}_{R'}(\tau;\lambda_{2}^{I},\lambda_{2}^{a},C_{2})  = \mathcal{W}_{R'}(\tau;\lambda_{2}^{I},\lambda_{2}^{a},C_{2})  T_{R}(C_{1}) \;,
\end{equation}
\begin{equation}
\mathcal{T}_{R}(\tau;\lambda_{1}^{I},\lambda_{1}^{a},C_{1}) \mathcal{W}_{R'}(\tau;\lambda_{2}^{I},\lambda_{2}^{a},C_{2})  = \mathcal{W}_{R'}(\tau;\lambda_{2}^{I},\lambda_{2}^{a},C_{2})  \mathcal{T}_{R}(\tau;\lambda_{1}^{I},\lambda_{1}^{a},C_{1}) \;.
\end{equation}
We may construct the common eigenstate of $\mathcal{T}_{R}(\tau;\lambda_{1}^{I},\lambda_{1}^{a},C_{1})$ and $ \mathcal{W}_{R'}(\tau;\lambda_{2}^{I},\lambda_{2}^{a},C_{2}) $.

Similar with the previous discussion, complete orthogonal basis of the Hilbert space $ \mathcal{H} $ can be selected as $ \{\vert \Lambda \rangle|\; \forall \;\Lambda\}  $. The loop group is given by 
 \begin{equation}
\mathcal{L} := \{  U_{n}T( C_{n-1})U_{n-1}\cdots T( C_{2})U_{2}T( C_{1})U_{1} |\; \forall\;U_{i}, \forall \;C_{i},\forall\; n\} \;,
\end{equation}
under which, $  \{\vert \Lambda \rangle|\; \forall \;\Lambda\}   $ can be decomposed into the equivalent classes. $  \{\vert \Lambda \rangle|\; \forall \;\Lambda\} =\cup_{\hat{\Lambda}}E(\hat{\Lambda})$, where 
 \begin{equation}
E(\hat{\Lambda}):=\{ L \vert \hat{\Lambda}\rangle\;|\; \forall\;L\in \mathcal{L}\} =\{  \vert L\hat{\Lambda}\rangle\;|\; \forall\;L\in \mathcal{L}\}
 \end{equation}
is the equivalent class generated by $ \mathcal{L} $ with $ \hat{\Lambda} $ an arbitrary element in it. The sub-Hilbert space $\mathcal{H}[E(\hat{\Lambda})]$ generated by $E(\hat{\Lambda})  $ is the eigenspace of $  \mathcal{W}_{R}(\tau;\lambda^{I},\lambda^{a},C)   $ with the eigenvalue $  \mathcal{W}_{R}(\hat{\Lambda};\tau;\lambda^{I},\lambda^{a},C)   $.

$ \mathcal{H} = \oplus_{\hat{\Lambda}} \; \mathcal{H} [E(\hat{\Lambda}) ]  $. Common eigenstate of $\mathcal{T}_{R}(\tau;\lambda^{I},\lambda^{a},C)  $ and $ \mathcal{W}_{R}(\tau;\lambda^{I},\lambda^{a},C)   $ can be constructed in each $\mathcal{H} [E(\hat{\Lambda}) ]    $. Let 
\begin{equation}
\vert D  \rangle_{(\Lambda',\Lambda)} =\int dL\; g^{-1}(LA) \vert L\Lambda' \rangle
\end{equation}
with $ \Lambda' = (A'_{i},\Phi^{I},\Psi^{a} )    $, $ \Lambda = (A_{i},\Phi^{I},\Psi^{a} )  $. $ \vert D  \rangle_{(\Lambda',\Lambda)}  \in  \mathcal{H} [E(\Lambda') ]   $. $U\vert D  \rangle_{(\Lambda',\Lambda)} = \vert D  \rangle_{(\Lambda',\Lambda)}  $, $ \vert D  \rangle_{(\Lambda',\Lambda)} \in \mathcal{H}_{ph} $.

The action of $\mathcal{T}_{R}(\tau;\lambda^{I},\lambda^{a},C)   $ on $ \vert D  \rangle_{(\Lambda',\Lambda)} $ is given by 
\begin{eqnarray}
\nonumber && \mathcal{T}_{R}(\tau;\lambda^{I},\lambda^{a},C) \vert D  \rangle_{(\Lambda',\Lambda)}\\ \nonumber &=& \frac{1}{d_{R}N_{(R,C)}}\int D_{R}U \;UT_{R}(C)W_{R}^{\Phi}(\tau;\lambda^{I},C)W_{R}^{\Psi}(\tau;\lambda^{a},C)\vert D  \rangle_{(\Lambda',\Lambda)}\\ &=&\nonumber
\frac{1}{d_{R}N_{(R,C)}}\int D_{R}U \;U\int dL \;W_{R}^{\Phi}(L\Phi^{I};\tau;\lambda^{I},C)W_{R}^{\Psi}(L\Psi^{a};\tau;\lambda^{a},C) g^{-1}(LA) \vert T^{-1}_{R}(C)L\Lambda'  \rangle
\\ &=&\nonumber
\frac{1}{d_{R}N_{(R,C)}}\int D_{R}U \;U\int dL \;W_{R}^{\Phi}(L\Phi^{I};\tau;\lambda^{I},C)W_{R}^{\Psi}(L\Psi^{a};\tau;\lambda^{a},C)  W_{R}^{A}(LA;C)g^{-1}(LA)\vert L\Lambda'    \rangle\\ &=&
\nonumber\frac{1}{d_{R}N_{(R,C)}}\int D_{R}U\int dL \;    W_{R}^{\Phi}(UL\Phi^{I};\tau;\lambda^{I},C)W_{R}^{\Psi}(UL\Psi^{a};\tau;\lambda^{a},C)  W_{R}^{A}(ULA;C)g^{-1}(LA) \vert L\Lambda'   \rangle\\ &=&\nonumber
\int dL \; \mathcal{W}_{R}(L\Lambda;\tau;\lambda^{I},\lambda^{a},C)  g^{-1}(LA) \vert L\Lambda'   \rangle\\ &=&
 \mathcal{W}_{R}(\Lambda;\tau;\lambda^{I},\lambda^{a},C) \vert D  \rangle_{(\Lambda',\Lambda)} \;.
\end{eqnarray} 
$\vert D  \rangle_{(\Lambda',\Lambda)}  $ is the state satisfying 
\begin{eqnarray}
\nonumber && \mathcal{T}_{R}(\tau;\lambda^{I},\lambda^{a},C) \vert D  \rangle_{(\Lambda',\Lambda)} = \mathcal{W}_{R}(\Lambda;\tau;\lambda^{I},\lambda^{a},C) \vert D  \rangle_{(\Lambda',\Lambda)} 
\\  && \mathcal{W}_{R}(\tau;\lambda^{I},\lambda^{a},C) \vert D  \rangle_{(\Lambda',\Lambda)} = \mathcal{W}_{R}(\Lambda';\tau;\lambda^{I},\lambda^{a},C) \vert D  \rangle_{(\Lambda',\Lambda)}  
\end{eqnarray} 
for the arbitrary $ R $, $\lambda^{I}$, $\lambda^{a}$ and $C$.

The action of $ \mathcal{T}_{R}^{+} $ is given by
\begin{equation}
\mathcal{T}_{R}^{+}(\tau;\lambda^{I},\lambda^{a},C) \vert D  \rangle_{(\Lambda',\Lambda)} = \mathcal{W}_{R}^{*}(\Lambda;\tau;\lambda^{I},\lambda^{a},C) \vert D  \rangle_{(\Lambda',\Lambda)} \;.
\end{equation}
So for $  \vert D  \rangle_{(\Lambda_{1},\Lambda_{2})}  $ and $  \vert D  \rangle_{(\Lambda'_{1},\Lambda'_{2})} $ with $  \Lambda_{1} = (A_{1i},\Phi^{I},\Psi^{a} )  $, $\Lambda_{2} = (A_{2i},\Phi^{I},\Psi^{a} )  $, $  \Lambda'_{1} = (A'_{1i},\Phi'^{I},\Psi'^{a} )  $, $\Lambda'_{2} = (A'_{2i},\Phi'^{I},\Psi'^{a} )  $, 
\begin{equation}
_{_{(\Lambda_{1},\Lambda_{2})}}\langle D | D \rangle_{_{(\Lambda'_{1},\Lambda'_{2})}}=0
\end{equation}
if $ \mathcal{W}_{R}(\Lambda_{1};\tau;\lambda^{I},\lambda^{a},C) \neq \mathcal{W}_{R}(\Lambda'_{1};\tau;\lambda^{I},\lambda^{a},C)  $ or $\mathcal{W}_{R}(\Lambda_{2};\tau;\lambda^{I},\lambda^{a},C) \neq \mathcal{W}_{R}(\Lambda'_{2};\tau;\lambda^{I},\lambda^{a},C)   $.

Moreover, in analogy with the discussion in Appendix \ref{BBB}, we have
\begin{equation}
\int DA\;g(LA)\vert D  \rangle_{(\Lambda',\Lambda)}=\vert L\Lambda'  \rangle_{ph}\;,
\end{equation}
so $ \{ \vert D  \rangle_{(\Lambda',\Lambda)}  |\; \forall \; A   \} $ forms the complete basis of $ \mathcal{H}_{ph}[E(\Lambda')]  $ and then, $ \{ \vert D  \rangle_{(\Lambda',\Lambda)}  |\; \forall \; \Lambda, \Lambda' \} $ composes the over-complete basis for $ \mathcal{H}_{ph}$. $ \vert D  \rangle_{(\Lambda',\Lambda)}=\vert D  \rangle_{(L\Lambda',L\Lambda)} $.

In $ U(1) $ case, 
\begin{equation}
\mathcal{T}(\tau;\lambda^{I},\lambda^{a},C)=T(\tau;\lambda^{I},\lambda^{a},C)=T(C)W^{\Phi}(\tau;\lambda^{I},C)W^{\Psi}(\tau;\lambda^{a},C)\;.
\end{equation}
The loop group $ \mathcal{L} $ only acts on $ A $, so $\vert D  \rangle_{(\Lambda',\Lambda)} = \vert D  \rangle_{(A',A)}  \vert \Phi , \Psi  \rangle  $ with $  \vert D  \rangle_{(A',A)}= \vert D  \rangle_{(\hat{A}',\hat{A})}  $ given by (\ref{u1u}).

\subsection{Modified supersymmetric 't Hooft operator }

There is no exact distinction between the spectrum of 
\begin{equation}
\mathcal{W}_{R}(\tau;\lambda^{I},\lambda^{a},C)=\dfrac{1}{d_{R}} tr P\exp \{i \oint_{C}\; (A_{R}^{i} dx_{i}+ (\frac{\tau_{2}}{2\pi})^{\frac{1}{2}} \Phi_{R}^{I} \lambda_{I} ds+(\frac{\tau_{2}}{2\pi})^{\frac{1}{2}}  \Psi_{R}^{a}\lambda_{a}ds ) \}
\end{equation}
and the spectrum of 
\begin{equation}
\mathcal{W}_{R}(C)=\dfrac{1}{d_{R}} tr P\exp \{i \oint_{C}\; (A_{R}^{i} dx_{i}) \}\;.
\end{equation}
Suppose $ y^{I} (x)$ and $ z^{a} (x)$ are functions specifying a three-dimensional hypersurface in superspace $(x^{i},y^{I},z^{a})$, for the special $ \lambda^{I} $ and $ \lambda^{a} $ with $\lambda^{I} = \partial_{i} y^{I}  \dot{x}^{i}$, $ \lambda^{a} = \partial_{i} z^{a}  \dot{x}^{i} $,
\begin{equation}
A_{R}^{i} dx_{i}+ (\frac{\tau_{2}}{2\pi})^{\frac{1}{2}} \Phi_{R}^{I} \lambda_{I} ds+(\frac{\tau_{2}}{2\pi})^{\frac{1}{2}}  \Psi_{R}^{a}\lambda_{a}ds  = A_{R}^{i} dx_{i}+ (\frac{\tau_{2}}{2\pi})^{\frac{1}{2}} \Phi_{R}^{I} dy_{I}+(\frac{\tau_{2}}{2\pi})^{\frac{1}{2}}  \Psi_{R}^{a}dz_{a}  \;,
\end{equation}
\begin{equation}
\mathcal{W}_{R}(\tau;\lambda^{I},\lambda^{a},C)=\dfrac{1}{d_{R}} tr P\exp \{i \oint_{C}ds\; [A_{R}^{i} + (\frac{\tau_{2}}{2\pi})^{\frac{1}{2}} \Phi_{R}^{I} \partial^{i}y_{I}+(\frac{\tau_{2}}{2\pi})^{\frac{1}{2}}  \Psi_{R}^{a}\partial^{i}z_{a}   ] \dot{x}_{i}\}\;.
\end{equation}
$\mathcal{W}_{R}(\tau;\lambda^{I},\lambda^{a},C)  $ could be taken as the Wilson loop of the gauge potential 
\begin{equation}
A_{R}^{' i}= A_{R}^{i} + (\frac{\tau_{2}}{2\pi})^{\frac{1}{2}}  \Phi_{R}^{I}  \partial^{i} y_{I}   +(\frac{\tau_{2}}{2\pi})^{\frac{1}{2}}   \Psi_{R}^{a}\partial^{i} z_{a}\;.
\end{equation}

Let
\begin{equation}
p(\tau;\Lambda; y,z )=\exp \{i \int d^{3}x \;(\frac{\tau_{2}}{2\pi})^{\frac{1}{2}}  tr [(\Phi_{I}  \partial_{i} y^{I}   + \Psi_{a}\partial_{i} z^{a})\Pi^{i}]\}\;,
\end{equation}
for $\lambda^{I} = \partial_{i} y^{I}  \dot{x}^{i}$, $ \lambda^{a} = \partial_{i} z^{a}  \dot{x}^{i} $, there will be  
\begin{equation}
p(\tau;\Lambda; y,z )W_{R}^{A}(C)p^{-1}(\tau;\Lambda; y,z )=W_{R}^{A}(C)W_{R}^{\Phi}(\tau;\lambda^{I},C)W_{R}^{\Psi}(\tau;\lambda^{a},C)\;
\end{equation}
and 
\begin{equation}
p(\tau;\Lambda; y,z )\mathcal{W}_{R}(C)p^{-1}(\tau;\Lambda; y,z )=\mathcal{W}_{R}(\tau;\lambda^{I},\lambda^{a},C)\;.
\end{equation}

The corresponding supersymmetric 't Hooft operator $ \mathcal{T}_{R}(\tau;\lambda^{I},\lambda^{a},C) $ and $ \mathcal{T}_{R}(C) $ can also be related by a unitary transformation. For operator 
\begin{equation}
q(\tau;\Lambda; y,z )=\exp \{-\frac{i}{2\pi}\int d^{3}x \;(\frac{\tau_{2}}{2\pi})^{\frac{1}{2}} tr [(\Phi_{I}  \partial_{i} y^{I}   + \Psi_{a}\partial_{i} z^{a})B^{i}]\}\;,
\end{equation}
\begin{equation}
T_{R} (C)q(\tau;\Lambda; y,z )T^{-1}_{R}(C)= \exp \{-\frac{i}{2\pi}\int d^{3}x \;(\frac{\tau_{2}}{2\pi})^{\frac{1}{2}} tr [H_{\vec{m}}(\Phi_{I}  \partial_{i} y^{I}   + \Psi_{a}\partial_{i} z^{a})b^{i}]\}q(\tau;\Lambda; y,z )\;,
\end{equation}
where we have used
\begin{equation}
T_{R} (C)B_{i}T^{-1}_{R}(C)=\Omega_{\vec{m}}^{-1}(\Sigma_{C})B_{i}\Omega_{\vec{m}}(\Sigma_{C})+H_{\vec{m}}b_{i}(C)\;.
\end{equation}
Therefore,
\begin{eqnarray}
\nonumber q(\tau;\Lambda; y,z )T_{R}(C)q^{-1}(\tau;\Lambda; y,z )&=& T_{R}(C) \exp \{\frac{i}{2\pi} \int d^{3}x \;(\frac{\tau_{2}}{2\pi})^{\frac{1}{2}} tr [H_{\vec{m}}(\Phi_{I}  \partial_{i} y^{I}   + \Psi_{a}\partial_{i} z^{a})b^{i}]\}\\\nonumber &=& T_{R}(C) \exp \{i \oint_{C} ds \;(\frac{\tau_{2}}{2\pi})^{\frac{1}{2}} 2tr [H_{\vec{m}}(\Phi_{I}  \partial_{i} y^{I}   + \Psi_{a}\partial_{i} z^{a})]\dot{x}^{i}\}
  \\ &=&
T_{R}(C) W_{R}^{\Phi}(\tau;\lambda^{I},C)W_{R}^{\Psi}(\tau;\lambda^{a},C)\;,
\end{eqnarray} 
and 
\begin{equation}
q(\tau;\Lambda; y,z )\mathcal{T}_{R}(C)q^{-1}(\tau;\Lambda; y,z )=\mathcal{T}_{R}(\tau;\lambda^{I},\lambda^{a},C)\;.
\end{equation}

More generically, with $ g(A) $ taken into account, suppose 
\begin{equation}
q(\tau;\Lambda; y,z,m )=\exp \{-\frac{i}{2\pi} \int d^{3}x \;(\frac{\tau_{2}}{2\pi})^{\frac{1}{2}} tr [(\Phi_{I}  \partial_{i} y^{I}   + \Psi_{a}\partial_{i} z^{a})B^{i}]\}g^{m}(A)\;,
\end{equation}
we will have 
\begin{equation}
q(\tau;\Lambda; y,z, m)T_{R}(C)q^{-1}(\tau;\Lambda; y,z ,m)=T_{R}(C)(W_{R}^{A} (C))^{m}W_{R}^{\Phi}(\tau;\lambda^{I},C)W_{R}^{\Psi}(\tau;\lambda^{a},C)
\end{equation}
and 
\begin{equation}
q(\tau;\Lambda; y,z, m)\mathcal{T}_{R}(C)q^{-1}(\tau;\Lambda; y,z ,m)=\mathcal{T}_{R}(\tau;m,\lambda^{I},\lambda^{a},C)\;,
\end{equation}
where
\begin{equation}
\mathcal{T}_{R}(\tau;m,\lambda^{I},\lambda^{a},C)=\frac{1}{d_{R} N_{(R,C)}}\int D_{R}U \; U[T_{R}(C)(W_{R}^{A} (C))^{m}W_{R}^{\Phi}(\tau;\lambda^{I},C)W_{R}^{\Psi}(\tau;\lambda^{a},C)]U^{-1}\;.
\end{equation}
This is the generalized T-transformation inserting a supersymmetric Wilson loop into the 't Hooft operator.

In path integral formulation, the obtained 't Hooft operator is 
\begin{equation}
\mathcal{T}'_{R}(\tau;\lambda^{I},\lambda^{a},C)=\frac{1}{d_{R}N_{(R,C)}}\int D_{R}U \; U[T_{R}(C)W_{R}^{\Phi}(\tau;\lambda^{I},C)W_{R}^{\Psi}(\tau;\lambda^{a},C)Y(\tau;\lambda^{I},\lambda^{a},C)]U^{-1}
\end{equation}
with $ Y(\tau;\lambda^{I},\lambda^{a},C) $ a unitary operator constructed from $ \Lambda $. It is expected that $  \mathcal{T}_{R}(\tau;\lambda^{I},\lambda^{a},C)$ and $ \mathcal{T}'_{R}(\tau;\lambda^{I},\lambda^{a},C) $ are equivalent.

\subsection{Mapping of the supersymmetric loop operators}

For the supersymmetric loop operators
\begin{equation}
\mathcal{T}_{R}(\tau;\lambda^{I},\lambda^{a},C)   =\frac{1}{d_{R}N_{(R,C)}}\int D_{R}U \;UT_{R}(C)\exp \{i \oint_{C} ds\;2tr[H_{\vec{m}}(\frac{\tau_{2}}{2\pi})^{\frac{1}{2}}(\Phi_{I}\lambda^{I}+\Psi_{a}\lambda^{a})]\}U^{-1}\;,
\end{equation}
\begin{equation}
\mathcal{W}_{R}(\tau;\lambda^{I},\lambda^{a},C)   =\frac{1}{d_{R}N_{(R,C)}}\int D_{R}U \;U\exp \{i \oint_{C}ds\; 2tr[H_{\vec{m}}(A_{i}\dot{x}^{i}+(\frac{\tau_{2}}{2\pi})^{\frac{1}{2}}(\Phi_{I}\lambda^{I}+\Psi_{a}\lambda^{a}))]\}U^{-1}\;, 
\end{equation}
we have got the basis $ \{ \vert D  \rangle_{(\Lambda',\Lambda)}  |\; \forall \; \Lambda, \Lambda' \} $ satisfying 
\begin{eqnarray}\label{Sl}
\nonumber  &&\mathcal{T}_{R}(\tau;\lambda^{I},\lambda^{a},C) \vert D  \rangle_{(\Lambda',\Lambda)}   =
\mathcal{W}_{R}(\Lambda;\tau;\lambda^{I},\lambda^{a},C) \vert D  \rangle_{(\Lambda',\Lambda)}   \;,
\\&&  \mathcal{W}_{R}(\tau;\lambda^{I},\lambda^{a},C) \vert D  \rangle_{(\Lambda',\Lambda)}    =
\mathcal{W}_{R}(\Lambda';\tau;\lambda^{I},\lambda^{a},C) \vert D  \rangle_{(\Lambda',\Lambda)}  
\;.
\end{eqnarray} 
It is possible to construct an $(R,\tau,\lambda^{I},\lambda^{a},C)$-independent unitary operator $ S_{1} $ with 
\begin{equation}\label{S}
S_{1}\mathcal{W}_{R}(\tau;\lambda^{I},\lambda^{a},C)S_{1}^{-1}=\mathcal{T}_{R}(\tau;\lambda^{I},\lambda^{a},C)\;,\;\;\;S_{1} \mathcal{T}_{R}(\tau;\lambda^{I},\lambda^{a},C)S_{1}^{-1}=\mathcal{W}^{+}_{R}(\tau;-\lambda^{I},-\lambda^{a},C)\;.
\end{equation}
The variation of $\lambda^{I}$ and $\lambda^{a}  $ gives
\begin{equation}\label{SS}
S_{1}\delta\mathcal{W}_{R}(\tau;\lambda^{I},\lambda^{a},C)S_{1}^{-1}=  \delta\mathcal{T}_{R}(\tau;\lambda^{I},\lambda^{a},C)\;,
\end{equation}
where 
\begin{eqnarray}
\nonumber  \delta \mathcal{W}_{R}(\tau;\lambda^{I},\lambda^{a},C)  &=& \frac{1}{d_{R}N_{(R,C)}}\int D_{R}U \;U\{i \oint_{C} ds\;2(\frac{\tau_{2}}{2\pi})^{\frac{1}{2}}tr[H_{\vec{m}}(\Phi_{I}\delta\lambda^{I}+\Psi_{a}\delta\lambda^{a})]\} \\ && W_{R}(\tau;\lambda^{I},\lambda^{a},C) U^{-1}\;,
\end{eqnarray} 
\begin{eqnarray}\label{SSSS}
\nonumber   \delta \mathcal{T}_{R}(\tau;\lambda^{I},\lambda^{a},C)  &=& \frac{1}{d_{R}N_{(R,C)}}\int D_{R}U \;U\{i \oint_{C}ds\;2(\frac{\tau_{2}}{2\pi})^{\frac{1}{2}} tr[H_{\vec{m}}(\Phi_{I}\delta\lambda^{I}+\Psi_{a}\delta\lambda^{a})]\} \\ && T_{R}(\tau;\lambda^{I},\lambda^{a},C) U^{-1}\;.
\end{eqnarray}

Again, (\ref{S}) cannot uniquely determine $ S_{1} $. From (\ref{Sl}) and (\ref{S}), $ S_{1}\vert D  \rangle_{(\Lambda',\Lambda)} $ is a state satisfying 
\begin{eqnarray}
\nonumber  &&\mathcal{T}_{R}(\tau;\lambda^{I},\lambda^{a},C)S_{1}\vert D  \rangle_{(\Lambda',\Lambda)}=\mathcal{W}_{R}(\Lambda';\tau;\lambda^{I},\lambda^{a},C)S_{1}\vert D  \rangle_{(\Lambda',\Lambda)}  
\\&& \mathcal{W}_{R}(\tau;\lambda^{I},\lambda^{a},C)S_{1}\vert D  \rangle_{(\Lambda',\Lambda)}=\mathcal{W}_{R}(\bar{\Lambda} ;\tau;\lambda^{I},\lambda^{a},C) S_{1} \vert D  \rangle_{(\Lambda',\Lambda)}\;,
\end{eqnarray}
where $ \bar{\Lambda}=(\bar{A}, \bar{\Phi},\bar{\Psi})=(-A^{M}  t^{M*} ,\Phi^{M}  t^{M*} ,\Psi^{M}  t^{M*} ) $. A trial solution  is
\begin{equation}\label{asz}
S_{1}\vert D  \rangle_{(\Lambda',\Lambda)}=\exp \{\frac{i}{4\pi}\int d^{3}x \; tr[(B_{i}-B'_{i})(A^{i}-A'^{i})]\}\vert D  \rangle^{*}_{(\bar{\Lambda},\bar{\Lambda}')}\;,
\end{equation}
where
\begin{equation}
\vert D  \rangle^{*}_{(\bar{\Lambda},\bar{\Lambda}')} =\int dL\; g(L\bar{A}') \vert L\bar{\Lambda} \rangle
\end{equation}
also composes the complete basis for $\mathcal{H}_{ph}  $. In $ U(1) $ situation, $ S_{1} $ is given by
\begin{eqnarray}\label{e123}
\nonumber && S_{1}B_{i}S_{1}^{-1}=2 \pi \Pi_{i}\;\;\;\;\;\;\;\;S_{1}\Phi^{I}S_{1}^{-1}=\Phi^{I}\;\;\;\;\;\;\;\;S_{1}\Psi^{a}S_{1}^{-1}=\Psi^{a}
\\ && S_{1}\Pi_{i}S_{1}^{-1}=-\frac{B_{i}}{2 \pi}\;\;\;\;\;\;\;\;S_{1}\Pi^{I}S_{1}^{-1}=\Pi^{I}\;\;\;\;\;\;\;\;S_{1}\Pi^{a}S_{1}^{-1}=\Pi^{a}\;,
\end{eqnarray} 
under which, (\ref{asz}) holds exactly.

In addition to $ S_{1} $, it is also necessary to introduce another unitary operator $ S_{2} $ inducing a rescaling: 
\begin{eqnarray}\label{e1234}
\nonumber && S_{2}A_{i}S_{2}^{-1}=A_{i}\;\;\;\;\;\;\;\;S_{2}\Phi^{I}S_{2}^{-1}=\frac{ \Phi^{I}}{|\tau|}\;\;\;\;\;\;\;\;S_{2}\Psi^{a}S_{2}^{-1}=\frac{ e^{\frac{i\theta}{2}}\Psi^{a}}{|\tau|}
\\ && S_{2}\Pi_{i}S_{2}^{-1}=\Pi_{i}\;\;\;\;\;\;\;\;S_{2}\Pi^{I}S_{2}^{-1}=\vert \tau\vert \Pi^{I}\;\;\;\;\;\;\;\;S_{2}\Pi^{a}S_{2}^{-1}=e^{-\frac{i\theta}{2}}|\tau|\Pi^{a}\;,
\end{eqnarray} 
$\tau = |\tau|e^{i\theta}$, $ \theta \in [0,\pi] $. The action of $ S_{2} $ on $\mathcal{W}  $ and $ \mathcal{T} $ is 
\begin{eqnarray}
 \nonumber && S_{2}\mathcal{W}_{R}(\tau;\lambda^{I},\lambda^{a},C)S_{2}^{-1}=\mathcal{W}_{R}(-\frac{1}{\tau};\lambda^{I},e^{\frac{i\theta}{2}}\lambda^{a},C)\;,
\\ && S_{2}\mathcal{T}_{R}(\tau;\lambda^{I},\lambda^{a},C)S_{2}^{-1}=\mathcal{T}_{R}(-\frac{1}{\tau};\lambda^{I},e^{\frac{i\theta}{2}}\lambda^{a},C)\;.
\end{eqnarray}

The S-transformation operator is $ S=S_{1} S_{2}$ with
\begin{eqnarray}\label{opl}
\nonumber  &&S\mathcal{W}_{R}(\tau;\lambda^{I},\lambda^{a},C)S^{-1}= \mathcal{T}_{R}(-\frac{1}{\tau};\lambda^{I},e^{\frac{i\theta}{2}}\lambda^{a},C)  \;,
\\ && S\mathcal{T}_{R}(\tau;\lambda^{I},\lambda^{a},C)S^{-1}=\mathcal{W}^{+}_{R}(-\frac{1}{\tau};-\lambda^{I},-e^{\frac{i\theta}{2}}\lambda^{a},C)\;.
\end{eqnarray}
This is the expected transformation rule for supersymmetric loop operators. When $ G=U(1) $, from (\ref{e123}) and (\ref{e1234}), 
\begin{eqnarray}\label{1qaz}
\nonumber && SB_{i}S^{-1}=2 \pi \Pi_{i}\;\;\;\;\;\;\;\;S\Phi^{I}S^{-1}=\frac{ \Phi^{I}  }{|\tau|}\;\;\;\;\;\;\;\;S\Psi^{a}S^{-1}=\frac{ e^{\frac{i\theta}{2}} \Psi^{a}    }{|\tau|}
\\ && S\Pi_{i}S^{-1}=-\frac{B_{i}}{2 \pi} \;\;\;\;\;\;\;\;S\Pi^{I}S^{-1}=\vert \tau\vert   \Pi^{I} \;\;\;\;\;\;\;\;S\Pi^{a}S^{-1}=e^{-\frac{i\theta}{2}}|\tau|     \Pi^{a}   \;,
\end{eqnarray} 
(\ref{opl}) is indeed satisfied.

$ S^{2}   $ is the charge conjugation operator with
\begin{eqnarray}
\nonumber  &&S^{2}\mathcal{W}_{R}(\tau;\lambda^{I},\lambda^{a},C)S^{-2}=\mathcal{W}^{+}_{R}(\tau;-\lambda^{I},-i\lambda^{a},C)  \;,
\\ && S^{2}\mathcal{T}_{R}(\tau;\lambda^{I},\lambda^{a},C)S^{-2}=\mathcal{T}^{+}_{R}(\tau;-\lambda^{I},-i\lambda^{a},C)\;.
\end{eqnarray}
Actions of $ S^{2} $ on $ \Lambda $ and $ \vert \Lambda \rangle $ are
\begin{eqnarray}\label{S2d}
\nonumber && S^{2}A^{M}_{i}t^{M}S^{-2}=-A^{M}_{i}t^{M*}\;\;\;\;\;S^{2}\Phi^{M}_{I}t^{M}S^{-2}=\Phi^{M}_{I}t^{M*}\;\;\;\;\;S^{2}\Psi^{M}_{a}t^{M}S^{-2}=i\Psi^{M}_{a}t^{M*}
\\ && S^{2}\Pi^{M}_{i}t^{M}S^{-2}=-\Pi^{M}_{i}t^{M*}\;\;\;\;\;S^{2}\Pi^{M}_{I}t^{M}S^{-2}= \Pi^{M}_{I}t^{M*}\;\;\;\;\;S^{2}\Pi^{M}_{a}t^{M}S^{-2}=-i\Pi^{M}_{a}t^{M*}\;.
\end{eqnarray}
and
\begin{equation}
S^{2}\vert \Lambda \rangle = \vert \Lambda^{C} \rangle\;,
\end{equation}
where $  \Lambda^{C}= (-A^{M}  t^{M*} ,\Phi^{M}  t^{M*} ,-i\Psi^{M}  t^{M*}  )   $ is the charge conjugate configuration of $ \Lambda   $, $ M=1,2,\cdots,\dim G $. 
\begin{equation}\label{opk}
\mathcal{W}_{R}(\Lambda^{C};\tau;\lambda^{I},\lambda^{a},C)=\mathcal{W}^{*}_{R}(\Lambda ;\tau;-\lambda^{I},-i\lambda^{a},C) \;.
\end{equation}

\subsection{Supersymmetry transformation of loop operators}

The definition of the loop operators and the S-duality transformation rules are both formulated at the kinematical level. In this subsection, we will begin to investigate the action of $ S $ at the dynamical level.

Suppose $ Q^{a}_{\alpha} $, $ \bar{Q}^{a}_{\dot{\alpha}} $, $S^{a\dot{\alpha}}  $, and $ \bar{S}^{a\alpha} $ with $a=1,2,3,4  $, $ \alpha, \dot{\alpha}=1,2$ are $ 32 $ supercharges of $ \mathcal{N} =4$ SYM theory, 
\begin{eqnarray}\label{sup111}
&&\nonumber Q^{a}_{\alpha}=\int d^{3}x\;J^{a}_{0\alpha}\;\;\;\;\;\;\;\;\;\bar{Q}_{a\dot{\alpha}}=\int d^{3}x\;\bar{J}_{0a\dot{\alpha}}\\ &&\nonumber S^{a\dot{\alpha}}=\int d^{3}x  \;(x^{\mu}\sigma_{\mu}^{\alpha\dot{\alpha}}J^{a}_{0\alpha}+\frac{\tau_{2}}{\pi}tr\{\Phi^{ab}\sigma^{0\alpha\dot{\alpha}}\Psi_{b\alpha}\})\;\;\;\;\;\;\;\;\;\bar{S}_{a}^{\alpha}=\int d^{3}x \;(x^{\mu}\sigma_{\mu}^{\alpha\dot{\alpha}}\bar{J}_{0a\dot{\alpha}}+2tr\{\Phi_{ab}\Pi^{b\alpha}\})\;,\\
\end{eqnarray}
where the supercharge density is 
\begin{eqnarray}\label{sup1}
\nonumber J^{a}_{0\beta}(\tau)&=&tr\{\frac{2\pi}{\tau_{2}}\Pi_{\alpha\beta}\Pi^{a\alpha}+\frac{\tau}{\tau_{2}} B_{\alpha\beta}\Pi^{a\alpha}+\Psi_{b \beta}\Pi^{ab}+\epsilon_{\alpha\beta}\Pi^{b\alpha}[\Phi_{bc},\Phi^{ac}]+\frac{\tau_{2}}{2 \pi}\Psi_{b\alpha}\sigma^{0\alpha \dot{\alpha}}\sigma^{i}_{\beta\dot{\alpha}}D_{i}\Phi^{ab}\}\\\nonumber \bar{J}_{0a\dot{\beta}}(\tau)&=&tr\{\Psi^{\alpha}_{a}\sigma_{\dot{\beta}}^{0\beta}\Pi_{\alpha\beta}+\frac{\bar{\tau}}{2 \pi}\Psi^{\alpha}_{a}\sigma_{\dot{\beta}}^{0\beta}B_{\alpha\beta}+\frac{2\pi}{\tau_{2}}\Pi^{b \beta}\sigma^{0}_{\beta\dot{\beta}}\Pi_{ab}-\frac{\tau_{2}}{2 \pi}\epsilon_{\dot{\alpha}\dot{\beta}}\Psi_{b\alpha}\sigma^{0\alpha\dot{\alpha}}[\Phi^{bc},\Phi_{ac}]\\ &&+\;\Pi^{b\alpha}\sigma^{i}_{\alpha\dot{\beta}}D_{i}\Phi_{ab}\}\;.
\end{eqnarray}
If \cite{2, 42, 456}
\begin{equation}\label{90}
 SQ^{a}_{\alpha}(\tau)S^{-1}=e^{\frac{i\theta}{2}}Q^{a}_{\alpha}(-\frac{1}{\tau})\;,\;\;\;\;\; S\bar{Q}_{a\dot{\alpha}}(\tau)S^{-1}=e^{-\frac{i\theta}{2}}\bar{Q}_{a\dot{\alpha}}(-\frac{1}{\tau})\;,
\end{equation}
\begin{equation}\label{909}
 S S^{a\dot{\alpha}}(\tau)S^{-1}=e^{\frac{i\theta}{2}}S^{a\dot{\alpha}}(-\frac{1}{\tau})\;,\;\;\;\;\; S\bar{S}^{a\alpha}(\tau)S^{-1}=e^{-\frac{i\theta}{2}}\bar{S}^{a\alpha}(-\frac{1}{\tau})\;,
\end{equation}
then according to the superconformal algebra, all of the superconformal charges will transform with the corresponding $ U(1)_{Y} $ phase and the theory will be S-duality invariant \cite{SHO}. When $ G=U(1) $, $ S $ is given by (\ref{1qaz}), (\ref{90}) and (\ref{909}) can be verified directly. In non-Abelian situation, we only have a definition of $ S $ through its action on loop operators, so in the following, we will study the supersymmetry variations of loop operators to give a primary check for (\ref{90}) and (\ref{909}).

The actions of $ W_{R}^{A} (C)$ and $ T_{R}(C) $ on supercharges both bring a loop integration:
\begin{eqnarray}
\nonumber &&W_{R}^{A}(C)\theta^{\beta}_{a}Q^{a}_{\beta}(\tau)W_{R}^{A-1}(C)=\theta^{\beta}_{a}Q^{a}_{\beta}(\tau)+\dfrac{4\pi}{\tau_{2}}\oint_{C}ds\; tr(H_{\vec{m}}\Pi^{a\alpha}) \theta^{\beta}_{a}  \dot{x}_{\alpha\beta}\\\nonumber &&
T_{R}(C)\theta^{\beta}_{a}Q^{a}_{\beta}(\tau)T_{R}^{-1}(C)=\theta^{\beta}_{a}Q^{a}_{\beta}(\tau)-\dfrac{4\pi \tau}{\tau_{2}}\oint_{C}
ds\; tr(H_{\vec{m}}\Pi^{a\alpha})\theta^{\beta}_{a}\dot{x}_{\alpha\beta}
\\\nonumber &&
W_{R}^{A}(C)\bar{\theta}^{a\dot{\beta}}\bar{Q}_{a\dot{\beta}}(\tau)W_{R}^{A-1}(C)=\bar{\theta}^{a\dot{\beta}}\bar{Q}_{a\dot{\beta}}(\tau)+2\oint_{C} ds\; tr(H_{\vec{m}}\Psi^{\alpha}_{a})\bar{\theta}^{a\dot{\beta}}\sigma^{0\beta}_{\dot{\beta}} \dot{x}_{\alpha\beta}
\\ && T_{R}(C)\bar{\theta}^{a\dot{\beta}}\bar{Q}_{a\dot{\beta}}(\tau)T_{R}^{-1}(C)=\bar{\theta}^{a\dot{\beta}}\bar{Q}_{a\dot{\beta}}(\tau)-2\bar{\tau}    \oint_{C}ds\;
   tr(H_{\vec{m}}  \Psi^{\alpha}_{a})\bar{\theta}^{a\dot{\beta}}\sigma^{0\beta}_{\dot{\beta}} \dot{x}_{\alpha\beta}\;.
\end{eqnarray}
\begin{eqnarray}
\nonumber &&W_{R}^{A}(C)\bar{\theta}_{a\dot{\beta}}S^{a\dot{\beta}}(\tau)W_{R}^{A-1}(C)=\bar{\theta}_{a\dot{\beta}}S^{a\dot{\beta}}(\tau)+\dfrac{4\pi}{\tau_{2}}\oint_{C}ds\; tr(H_{\vec{m}}\Pi^{a\alpha}) x^{\mu}\sigma_{\mu}^{\beta\dot{\beta}}\bar{\theta}_{a\dot{\beta}}  \dot{x}_{\alpha\beta}\\\nonumber &&
T_{R}(C)\bar{\theta}_{a\dot{\beta}}S^{a\dot{\beta}}(\tau)T_{R}^{-1}(C)=\bar{\theta}_{a\dot{\beta}}S^{a\dot{\beta}}(\tau)-\dfrac{4\pi \tau}{\tau_{2}}\oint_{C}
ds\; tr(H_{\vec{m}}\Pi^{a\alpha})x^{\mu}\sigma_{\mu}^{\beta\dot{\beta}}\bar{\theta}_{a\dot{\beta}}\dot{x}_{\alpha\beta}
\\\nonumber &&
W_{R}^{A}(C)\theta^{a}_{\beta}\bar{S}_{a}^{\beta}(\tau)W_{R}^{A-1}(C)=\theta^{a}_{\beta}\bar{S}_{a}^{\beta}(\tau)+2\oint_{C} ds\; tr(H_{\vec{m}}\Psi^{\alpha}_{a})x^{\mu}\sigma_{\mu}^{\beta\dot{\beta}}\theta^{a}_{\beta}\sigma^{0\beta}_{\dot{\beta}} \dot{x}_{\alpha\beta}
\\ && T_{R}(C)\theta^{a}_{\beta}\bar{S}_{a}^{\beta}(\tau)T_{R}^{-1}(C)=\theta^{a}_{\beta}\bar{S}_{a}^{\beta}(\tau)-2\bar{\tau}    \oint_{C}ds\;
   tr(H_{\vec{m}}  \Psi^{\alpha}_{a})x^{\mu}\sigma_{\mu}^{\beta\dot{\beta}}\theta^{a}_{\beta}\sigma^{0\beta}_{\dot{\beta}} \dot{x}_{\alpha\beta}\;.
\end{eqnarray}
Consider the supersymmetry variations of the loop operators and for simplicity, suppose $ \lambda^{a} =0$, then 
\begin{eqnarray}
\nonumber&& W_{R}(\tau;\lambda^{ab},C) =W^{A}_{R}(C) W_{R}^{\Phi}(\tau;\lambda^{ab},C)  =\exp \{i \oint_{C}ds\; 2tr[H_{\vec{m}}(A_{\alpha\beta}\dot{x}^{\alpha\beta}+(\frac{\tau_{2}}{2\pi})^{\frac{1}{2}}\Phi_{ab}\lambda^{ab})]\}\;,\\ && \mathcal{W}_{R}(\tau;\lambda^{ab},C) = \frac{1}{d_{R}N_{(R,C)}}\int D_{R}U \;UW_{R}(\tau;\lambda^{ab},C)U^{-1}\;,
\end{eqnarray}
\begin{eqnarray}
\nonumber&& T_{R}(\tau;\lambda^{ab},C)=T_{R}(C) W_{R}^{\Phi}(\tau;\lambda^{ab},C)=T_{R}(C)\exp \{i \oint_{C}ds\; 2tr[H_{\vec{m}}(\frac{\tau_{2}}{2\pi})^{\frac{1}{2}}\Phi_{ab}\lambda^{ab}]\}\;,\\ && \mathcal{T}_{R}(\tau;\lambda^{ab},C) = \frac{1}{d_{R}N_{(R,C)}}\int D_{R}U \;UT_{R}(\tau;\lambda^{ab},C)U^{-1}\;.
\end{eqnarray}
Direct calculation gives 
\begin{eqnarray}\label{laq1}
\nonumber&& [\theta^{\beta}_{a}Q^{a}_{\beta}(\tau),T_{R}(C)]=[ \dfrac{4\pi \tau}{\tau_{2}} \oint_{C}ds\;
  tr(H_{\vec{m}}\Pi^{a\alpha})\theta^{\beta}_{a} \dot{x}_{\alpha\beta}]T_{R}(C)\;\\ && [\bar{\theta}^{a\dot{\beta}}\bar{Q}_{a\dot{\beta}}(\tau),T_{R}(C)]=[2\bar{\tau}    \oint_{C}ds\;
   tr(H_{\vec{m}}  \Psi^{\alpha}_{a})\bar{\theta}^{a\dot{\beta}}\sigma^{0\beta}_{\dot{\beta}} \dot{x}_{\alpha\beta}]T_{R}(C)\;
\end{eqnarray}
\begin{eqnarray}\label{laq11}
\nonumber&& [\theta^{\beta}_{a}Q^{a}_{\beta}(\tau),W_{R}^{A}(C)]=-[\dfrac{4\pi}{\tau_{2}}\oint_{C}ds\;  tr(H_{\vec{m}}\Pi^{a\alpha})\theta^{\beta}_{a}\dot{x}_{\alpha\beta}]W_{R}^{A}(C)\;\\ && [\bar{\theta}^{a\dot{\beta}}\bar{Q}_{a\dot{\beta}}(\tau),W_{R}^{A}(C)]=-[2 \oint_{C} ds\; tr(H_{\vec{m}}\Psi^{\alpha}_{a})\bar{\theta}^{a\dot{\beta}}\sigma^{0\beta}_{\dot{\beta}} \dot{x}_{\alpha\beta}]W_{R}^{A}(C)\;
\end{eqnarray}
\begin{eqnarray}\label{laq111}
\nonumber&& [\theta^{\beta}_{a}Q^{a}_{\beta}(\tau),W_{R}^{\Phi}(\tau;\lambda^{ab},C) ]=-[2(\frac{\tau_{2}}{2\pi})^{\frac{1}{2}}\oint_{C}ds\; tr(H_{\vec{m}}\Psi^{a}_{\alpha})\theta^{b\alpha}\lambda_{ab}]W_{R}^{\Phi}(\tau;\lambda^{ab},C) \;\\\nonumber && [\bar{\theta}^{a\dot{\beta}}\bar{Q}_{a\dot{\beta}}(\tau),W_{R}^{\Phi}(\tau;\lambda^{ab},C) ]=-[2(\frac{\tau_{2}}{2\pi})^{-\frac{1}{2}} \oint_{C} ds\; tr(H_{\vec{m}}\Pi^{a\alpha})\sigma^{0}_{\alpha\dot{\alpha}}\bar{\theta}^{b\dot{\alpha}}\lambda_{ab}]W_{R}^{\Phi}(\tau;\lambda^{ab},C) \;.\\
\end{eqnarray}
So the supersymmetry variations of $\mathcal{W}_{R}(\tau;\lambda^{ab},C)  $ and $ \mathcal{T}_{R}(\tau;\lambda^{ab},C)   $ are 
\begin{eqnarray}\label{pl}
&&\nonumber  [\theta^{\beta}_{a}Q^{a}_{\beta}(\tau)+\bar{\theta}^{a\dot{\beta}}\bar{Q}_{a\dot{\beta}}(\tau),\mathcal{W}_{R}(\tau;\lambda^{ab},C) ]\\&=&\nonumber  \frac{1}{d_{R}N_{(R,C)}}\int D_{R}U \;U[\oint_{C}ds\;  tr\{H_{\vec{m}}[\Pi^{a\alpha}(-(\frac{4\pi}{\tau_{2}})\theta^{\beta}_{a}\dot{x}_{\alpha\beta}-2(\frac{\tau_{2}}{2\pi})^{-\frac{1}{2}}\sigma^{0}_{\alpha\dot{\alpha}}\bar{\theta}^{b\dot{\alpha}}\lambda_{ab})\\ &+&  \Psi^{\alpha}_{a}(-2\sigma^{0\beta}_{\dot{\beta}}\bar{\theta}^{a\dot{\beta}}\dot{x}_{\alpha\beta}-2(\frac{\tau_{2}}{2\pi})^{\frac{1}{2}}\theta_{b\alpha}\lambda^{ab})]\}]W_{R}(\tau;\lambda^{ab},C)U^{-1}  \;,
\end{eqnarray}
\begin{eqnarray}\label{pll}
&&\nonumber [\theta^{\beta}_{a}Q^{a}_{\beta}(\tau)+\bar{\theta}^{a\dot{\beta}}\bar{Q}_{a\dot{\beta}}(\tau),\mathcal{T}_{R}(\tau;\lambda^{ab},C) ]\\&=&\nonumber  \frac{1}{d_{R}N_{(R,C)}}\int D_{R}U \;U [\oint_{C}ds\; tr\{ H_{\vec{m}}[\Pi^{a\alpha}((\frac{4\pi\tau}{\tau_{2}}) \dot{x}_{\alpha\beta} \theta^{\beta}_{a}-2(\frac{\tau_{2}}{2\pi})^{-\frac{1}{2}}\sigma^{0}_{\alpha\dot{\alpha}}\bar{\theta}^{b\dot{\alpha}}\lambda_{ab})\\ &+&  \Psi^{\alpha}_{a}( 2\bar{\tau} \bar{\theta}^{a\dot{\beta}}\sigma^{0\beta}_{\dot{\beta}} \dot{x}_{\alpha\beta}-2(\frac{\tau_{2}}{2\pi})^{\frac{1}{2}}\theta_{b\alpha}\lambda^{ab}) ]\}]T_{R}(\tau;\lambda^{ab},C) U^{-1}  \;.
\end{eqnarray}
Similar conclusions hold for $ \bar{\theta}_{a\dot{\beta}}S^{a\dot{\beta}} $ and $ \theta^{a}_{\beta}\bar{S}_{a}^{\beta} $ with $( \theta^{\beta}_{a} ,\bar{\theta}^{a\dot{\beta}} ) $ replaced by $(x^{\mu}\sigma_{\mu}^{\beta\dot{\beta}}\bar{\theta}_{a\dot{\beta}}  , x^{\mu}\sigma_{\mu}^{\beta\dot{\beta}}\theta^{a}_{\beta} )$ on the right-hand sides of (\ref{pl}) and (\ref{pll}).

Under the action of the S-transformation operator $ S=S_{1} S_{2}$, (\ref{pl}) becomes 
\begin{eqnarray}\label{pl1}
&&\nonumber   [S\theta^{\beta}_{a}Q^{a}_{\beta}(\tau)S^{-1}+S\bar{\theta}^{a\dot{\beta}}\bar{Q}_{a\dot{\beta}}(\tau)S^{-1},S\mathcal{W}_{R}(\tau;\lambda^{ab},C)S^{-1} ]\\&=&\nonumber \frac{1}{d_{R}N_{(R,C)}} S_{1}\int D_{R}U \;U\oint_{C}ds\;  tr\{H_{\vec{m}}[\Pi^{a\alpha}e^{-\frac{i\theta}{2}}|\tau|(-(\frac{4\pi  }{\tau_{2}})\theta^{\beta}_{a}\dot{x}_{\alpha\beta}-2  (\frac{\tau_{2}}{2\pi})^{-\frac{1}{2}}\sigma^{0}_{\alpha\dot{\alpha}}\bar{\theta}^{b\dot{\alpha}}\lambda_{ab})\\\nonumber  &+& \frac{ e^{\frac{i\theta}{2}}\Psi^{\alpha}_{a}}{|\tau|} (-2\sigma^{0\beta}_{\dot{\beta}}\bar{\theta}^{a\dot{\beta}}\dot{x}_{\alpha\beta}-2(\frac{\tau_{2}}{2\pi})^{\frac{1}{2}}\theta_{b\alpha}\lambda^{ab})]\}W_{R}(-\frac{1}{\tau};\lambda^{ab},C)U^{-1} S_{1}^{-1}\\ \nonumber &=& \frac{1}{d_{R}N_{(R,C)}}S_{1} \int D_{R}U \;U\oint_{C}ds\; tr\{H_{\vec{m}}[\Pi^{a\alpha} \delta \lambda'_{a\alpha}+  \Psi^{\alpha}_{a}\delta \lambda_{\alpha}^{a}]\}W_{R}(-\frac{1}{\tau};\lambda^{ab},C)U^{-1} S_{1}^{-1} \;,\\
\end{eqnarray}
where 
\begin{eqnarray}
&&\delta \lambda'_{a\alpha} =e^{-\frac{i\theta}{2}}|\tau|(-(\frac{4\pi  }{\tau_{2}})\theta^{\beta}_{a}\dot{x}_{\alpha\beta}-2  (\frac{\tau_{2}}{2\pi})^{-\frac{1}{2}}\sigma^{0}_{\alpha\dot{\alpha}}\bar{\theta}^{b\dot{\alpha}}\lambda_{ab}) \;,\\ &&\delta \lambda_{\alpha}^{a}=\frac{ e^{\frac{i\theta}{2}}}{|\tau|} (-2\sigma^{0\beta}_{\dot{\beta}}\bar{\theta}^{a\dot{\beta}}\dot{x}_{\alpha\beta}-2(\frac{\tau_{2}}{2\pi})^{\frac{1}{2}}\theta_{b\alpha}\lambda^{ab})\;.
\end{eqnarray}
On the other hand, with $ \tau $ replaced by $ -1/\tau $, and $\theta^{\beta}_{a}  $, $ \bar{\theta}^{a\dot{\beta}} $ replaced by $e^{\frac{i\theta}{2}}\theta^{\beta}_{a}  $, $e^{-\frac{i\theta}{2}} \bar{\theta}^{a\dot{\beta}} $, (\ref{pll}) becomes 
\begin{eqnarray}\label{pl2}
&&\nonumber  [e^{\frac{i\theta}{2}}\theta^{\beta}_{a}Q^{a}_{\beta}(-\frac{1}{\tau})+e^{-\frac{i\theta}{2}}\bar{\theta}^{a\dot{\beta}}\bar{Q}_{a\dot{\beta}}(-\frac{1}{\tau}),\mathcal{T}_{R}(-\frac{1}{\tau};\lambda^{ab},C) ] \\&=& \frac{1}{d_{R}N_{(R,C)}}\int D_{R}U \;U [\oint_{C}ds\; tr\{H_{\vec{m}}[\Pi^{a\alpha}\delta \lambda'_{a\alpha}+ \Psi^{\alpha}_{a}\delta \lambda_{\alpha}^{a} ]\}]T_{R}(-\frac{1}{\tau};\lambda^{ab},C) U^{-1}  \;.
\end{eqnarray}
According to (\ref{SS})-(\ref{SSSS}), the right-hand sides of (\ref{pl1}) and (\ref{pl2}) are equal. In other words, for
\begin{equation}\label{Delta}
\Delta^{a}_{\beta}(\tau):= SQ^{a}_{\beta}(\tau)S^{-1}-e^{\frac{i\theta}{2}}Q^{a}_{\beta}(-\frac{1}{\tau})\;,\;\;\;\;\;\bar{\Delta}_{a\dot{\beta}}(\tau):= S\bar{Q}_{a\dot{\beta}}(\tau)S^{-1}-e^{-\frac{i\theta}{2}}\bar{Q}_{a\dot{\beta}}(-\frac{1}{\tau})\;,
\end{equation}
we have 
\begin{equation}\label{89}
 [\theta^{\beta}_{a}\Delta^{a}_{\beta}(\tau)+\bar{\theta}^{a\dot{\beta}}\bar{\Delta}_{a\dot{\beta}}(\tau),\mathcal{T}_{R}(-\frac{1}{\tau};\lambda^{ab},C)]=0
\end{equation}
or equivalently, 
\begin{equation}\label{88}
 [\theta^{\beta}_{a}S^{-1}\Delta^{a}_{\beta}(\tau)S+\bar{\theta}^{a\dot{\beta}}S^{-1}\bar{\Delta}_{a\dot{\beta}}(\tau)S,\mathcal{W}_{R}(\tau;\lambda^{ab},C)]=0
\end{equation}
for the arbitrary $ \theta^{\beta}_{a}, \bar{\theta}^{a\dot{\beta}}, \lambda^{ab}$, and $C $. The simplest, but not the only, possibility is $\Delta^{a}_{\beta}(\tau)=\bar{\Delta}_{a\dot{\beta}}(\tau)=0    $ which is just (\ref{90}). In (\ref{Delta}), with $Q^{a}_{\alpha}  $ and $ \bar{Q}_{a\dot{\alpha}} $ replaced by $S^{a\dot{\alpha}}  $ and $ \bar{S}^{a\alpha} $, we will have the similar result for (\ref{909}).

When $ \theta^{\beta}_{a}\dot{x}_{\alpha\beta}+  (\frac{\tau_{2}}{2\pi})^{\frac{1}{2}}\sigma^{0}_{\alpha\dot{\alpha}}\bar{\theta}^{b\dot{\alpha}}\lambda_{ab}=0 $, $\delta \lambda'_{a\alpha} =\delta \lambda_{\alpha}^{a}=0  $,
\begin{eqnarray}
&& [\theta^{\beta}_{a}Q^{a}_{\beta}(\tau)+\bar{\theta}^{a\dot{\beta}}\bar{Q}_{a\dot{\beta}}(\tau),\mathcal{W}_{R}(\tau;\lambda^{ab},C) ]=0\;,\\ && [e^{\frac{i\theta}{2}}\theta^{\beta}_{a}Q^{a}_{\beta}(-\frac{1}{\tau})+e^{-\frac{i\theta}{2}}\bar{\theta}^{a\dot{\beta}}\bar{Q}_{a\dot{\beta}}(-\frac{1}{\tau}),\mathcal{T}_{R}(-\frac{1}{\tau};\lambda^{ab},C) ]=0\;.
\end{eqnarray}
Similarly, when the special superconformal symmetry is preserved, 
\begin{eqnarray}
&& [\bar{\theta}_{a\dot{\beta}}S^{a\dot{\beta}}(\tau)+\theta_{a\beta}\bar{S}^{a\beta}(\tau),\mathcal{W}_{R}(\tau;\lambda^{ab},C) ]=0\;,\\ && [e^{\frac{i\theta}{2}}\bar{\theta}_{a\dot{\beta}}S^{a\dot{\beta}}(-\frac{1}{\tau})+e^{-\frac{i\theta}{2}}\theta_{a\beta}\bar{S}^{a\beta}(-\frac{1}{\tau}),\mathcal{T}_{R}(-\frac{1}{\tau};\lambda^{ab},C) ]=0\;.
\end{eqnarray}
The supersymmetry preserved by $\mathcal{W}_{R}(\tau;\lambda^{ab},C)  $ in theory with the coupling constant $ \tau $ and those preserved by $\mathcal{T}_{R}(-1/\tau;\lambda^{ab},C)  $ in theory with the coupling constant $ -1/\tau $ is related by a $ U(1)_{Y} $ phase, as is already shown in path integral formalism \cite{pa}.

From (\ref{90}) and (\ref{909}), 
\begin{equation}\label{90u}
 S^{2}Q^{a}_{\beta}(\tau)S^{-2}=i Q^{a}_{\beta}(\tau)\;,\;\;\;\;\; S^{2}\bar{Q}_{a\dot{\beta}}(\tau)S^{-2}=-i\bar{Q}_{a\dot{\beta}}(\tau)
\end{equation}
\begin{equation}\label{90uj}
 S^{2}S^{a\dot{\beta}}(\tau)S^{-2}=i S^{a\dot{\beta}}(\tau)\;,\;\;\;\;\; S^{2}\bar{S}^{a\beta}(\tau)S^{-2}=-i\bar{S}^{a\beta}(\tau)\;.
\end{equation}
With (\ref{S2d}) plugged in (\ref{sup1}) and (\ref{sup111}), we have 
\begin{equation}
 S^{2}J^{a}_{0\beta}(\tau)S^{-2}=i J^{a}_{0\beta}(\tau)\;,\;\;\;\;\; S^{2}\bar{J}_{0a\dot{\beta}}(\tau)S^{-2}=-i\bar{J}_{0a\dot{\beta}}(\tau)\;,
\end{equation}
so (\ref{90u}) and (\ref{90uj}) are indeed satisfied.

With $ \mathcal{W}_{R}(\tau;\lambda^{I},\lambda^{a},C) $ and $ \mathcal{T}_{R}(\tau;\lambda^{I},\lambda^{a},C) $ given, it is straightforward to calculate the successive action of supercharges to get the whole supermultiplet. In addition to (\ref{laq1})-(\ref{laq111}), the supersymmetry variation of the fermionic Wilson loop 
\begin{equation}
 W_{R}^{\Psi}(\tau;\lambda^{a},C)=\exp \{i (\frac{\tau_{2}}{2\pi})^{\frac{1}{2}}\oint_{C}ds\;2 tr(H_{\vec{m}}\Psi_{a} )\lambda^{a} \}
\end{equation}
is given by
\begin{eqnarray}\label{law21}
\nonumber&&[ \theta^{\beta}_{a}Q^{a}_{\beta}(\tau),W_{R}^{\Psi}(\tau;\lambda^{a},C)]\\\nonumber &=& (\frac{\tau_{2}}{2\pi})^{\frac{1}{2}}\{\oint_{C}ds\; 2tr[H_{\vec{m}}(\frac{2\pi}{\tau_{2}}\Pi_{\alpha\beta}\lambda^{a\alpha}\theta^{\beta}_{a}+\frac{\tau}{\tau_{2}} B_{\alpha\beta}\lambda^{a\alpha}\theta^{\beta}_{a}+\lambda^{b\alpha}\theta^{\beta}_{a}\epsilon_{\alpha\beta}[\Phi_{bc},\Phi^{ac}]) ]\}W_{R}^{\Psi}(\tau;\lambda^{a},C) \\ \nonumber && [\bar{\theta}^{a\dot{\beta}}\bar{Q}_{a\dot{\beta}}(\tau),W_{R}^{\Psi}(\tau;\lambda^{a},C)]\\ &=& (\frac{\tau_{2}}{2\pi})^{\frac{1}{2}}\{\oint_{C}ds\;2 tr[H_{\vec{m}}(\lambda^{b\alpha}\bar{\theta}^{a\dot{\beta}} \sigma^{i}_{\alpha\dot{\beta}}D_{i}\Phi_{ab}+\frac{2\pi}{\tau_{2}}\lambda^{b\beta}\bar{\theta}^{a\dot{\beta}} \sigma^{0}_{\beta\dot{\beta}}\Pi_{ab}) ]\}W_{R}^{\Psi}(\tau;\lambda^{a},C) \;.
\end{eqnarray}
Equations (\ref{laq1})-(\ref{laq111}) and (\ref{law21}) compose the basic elements, from which, the arbitrary supersymmetry descendants $ [Q,\cdots [Q,\mathcal{W}_{R}  ] ] $ and $ [Q,\cdots [Q,\mathcal{T}_{R}  ] ] $ can be obtained. The descendant operators for $\mathcal{W}_{R}  $ and $ \mathcal{T}_{R} $ generically take the form of 
\begin{equation}\label{1q2}
 \frac{1}{d_{R}N_{(R,C)}}\int D_{R}U \;U[\oint_{C}ds \cdots +\oint_{C}ds\oint_{C}ds'\cdots+\cdots ]W_{R}(\tau;\lambda^{ab},\lambda^{a},C)U^{-1} 
\end{equation}
and 
\begin{equation}\label{1q21}
 \frac{1}{d_{R}N_{(R,C)}}\int D_{R}U \;U[\oint_{C}ds \cdots +\oint_{C}ds\oint_{C}ds'\cdots+\cdots ]T_{R}(\tau;\lambda^{ab},\lambda^{a},C)U^{-1} \;.
\end{equation}
Aside from $ A_{\alpha\beta},\Phi_{ab} ,\Psi_{a}$, the conjugate momentum $\Pi_{\alpha\beta},\Pi_{ab} , \Pi_{a}$ are also involved in ``$ \cdots $", which, in path integral formalism, can be replaced by the time derivatives of the fundamental fields. Equation (\ref{1q21}) also suggests the path integral definition of the 't Hooft descendants. Suppose $ \Lambda_{0} $ is the singular background related with $T_{R}  $ breaking the gauge symmetry into its invariant group, roughly, the path integral formulation of (\ref{1q21}) could be $\int D\Lambda\;  e^{i S(\Lambda_{0}+\Lambda) } [\cdots] $ with $ [\cdots]   $ borrowed from (\ref{1q21}) which is invariant under the residue gauge symmetry. Finally, the integration over $\{u \Lambda_{0}u^{-1} |\; \forall \; u\}$ should be carried out to restore the original gauge symmetry.

\section{Discussion}\label{discussion}

In this paper, we have constructed the gauge invariant 't Hooft operator in canonical formalism and studied its T-transformation and S-transformation rules. T-transformation is known to be realized via a unitary operator $ g(A) $ \cite{T}, under which, 't Hooft operators are found to transform as (\ref{twq1}), verifying the statement in \cite{10}. S-transformation is expected to be realized by a unitary operator $ S $ whose explicit form is unknown. The original aim of this paper is to reconstruct $ S $ through its action on loop operators. Especially, it is hoped that (\ref{2s}) could get the extension in YM theory as well as the $ \mathcal{N} =4$ SYM theory. We have demonstrated that Wilson and 't Hooft operators share the same spectrum thus could indeed be related by a unitary operator $ S $, which is not so obvious in non-Abelian theories. The mapping of loop operators under the S-transformation puts a strong constraint on $ S $, from which, a tentative $ S $ is constructed. However, just as $ e^{iX} $ and $ e^{2 \pi i P} $ in $ 1d $ quantum mechanics, Wilson and 't Hooft operators are highly degenerate thus could not act as the fundamental observables. We need to go further to study the mapping of flux operators to make $ S $ entirely determined.

For $ \mathcal{N} =4$ SYM theory, with $ S $ given, if the supercharges transform with a $ U(1)_{Y} $ factor, the theory will be S-duality invariant. We compute the supersymmetry variations of the loop operators with the fermionic couplings turned off and get the evidence for the $ U(1)_{Y} $ transformation of the supercharges, but the exact proof still needs the adequate knowledge of $ S $.

With $ \mathcal{W}_{R} $ and $ \mathcal{T}_{R} $ given, it is easy to get the whole supermultiplet (\ref{1q2}) and (\ref{1q21}). A byproduct of this paper is the explicit construction of the supersymmetry descendants for 't Hooft operators, which are difficult to define in path integral formalism. Classification of the superconformal defect multiplets and especially, the superconformal lines was recently studied in \cite{cite}. Our discussion is carried out in Minkowski spacetime and could be converted to Euclidean spacetime by a Wick rotation. Then when $ C $ is a circle in space, we can write down the $ 1/2 $ BPS Wilson and 't Hooft operators. Successive action of the supercharges gives the concrete operator definition of the whole $ 1/2 $ BPS multiplet of the defects.

\section*{Acknowledgments}

The work is supported in part by NSFC under Grant No. 11605049.


\begin{appendix}

\section{A direct proof for the path integral representation of the Wilson loop}\label{AAA}

When the gauge group is $ SU(N) $ and $ R $ is an irreducible representation, consider the path-ordered integration $P_{R}(A;C;t)  $ along the loop $ C $ parametrized by $ s \in [0,1) $:
\begin{equation}
P_{R}(A;C;t)=  P\exp \{i \int_{0}^{t}ds\; A_{R}^{i} \dot{x}_{i} \}=  P\exp \{i \int_{0}^{t}ds\; A_{R} \}\;,
\end{equation}
$P_{R}(A;C;0)=I$, $ \mathcal{W}_{R}(A;C)=\frac{1}{d_{R}} tr P_{R}(A;C;1) $.
\begin{equation}
\dfrac{d}{dt}P_{R}(A;C;t)=i A_{R} (t)P_{R}(A;C;t)\;.
\end{equation}
One can always find a gauge transformation $ U_{R} $, under which the original gauge field $ A_{R}^{i} $ becomes
\begin{equation}
A'^{i}_{R}=U^{-1}_{R}A_{R}^{i}U_{R}+iU^{-1}_{R}\partial^{i}U_{R}\;,
\end{equation}
where $ A'^{i}_{R} $ is a gauge field with $ A'_{R}=A'^{i}_{R}\dot{x}_{i}= H_{R}(A;C;1) $ a constant in Cartan subalgebra along the loop $ C $. 
\begin{equation}
U^{-1}_{R} P_{R}(A;C;1) U_{R}= \exp\{i \int_{0}^{1}ds\; H_{R}(A;C;1)\}= \exp\{i H_{R}(A;C;1)\}\;.
\end{equation}
So to calculate the Wilson loop, it is enough to consider the constant $A_{R}  $ in Cartan subalgebra at $ C $.

For definiteness, suppose $G= SU(2) $, for the representation $ R $ with the spin $ j=\dfrac{1}{2},1,\cdots $, from (\ref{ba}), 
\begin{eqnarray}\label{a1}
\nonumber  \mathcal{W}_{j}(A;C)  &=&\frac{1}{(2j+1)n_{(j,C)}} \int D_{j}u(s)   \;\exp \{i \int_{0}^{1}ds\; j\;tr[\sigma_{3} (u^{-1}A_{i}u+iu^{-1}\partial_{i}u) ]\dot{x}^{i} \} \\&=&\frac{1}{(2j+1)n_{(j,C)}}  \int D_{j}u(s)   \;\exp \{i \int_{0}^{1}ds\; j \;tr[\sigma_{3} (\frac{1}{2}au^{-1}\sigma_{3}u+iu^{-1}\dot{u}) ] \}\;,
\end{eqnarray}
where $a$ is a constant. On the other hand, 
\begin{equation}\label{a11}
\mathcal{W}_{j}(A;C)=\frac{1}{2j+1}tr \exp\{i aH_{j}\} =\frac{1}{2j+1}\sum^{j}_{m=-j} e^{i am}\;,
\end{equation}
it remains to show the right-hand sides of (\ref{a1}) and (\ref{a11}) are equal. The proof is in \cite{prof}. Here, we will give a review with a clarification on the concrete form of $ D_{j}u(s)  $ in (\ref{a1}).

The $ SU(2) $ matrix $ u $ can be parametrized by angles $ (\alpha,\beta,\gamma)$:
\begin{eqnarray}
\nonumber   u &=&\exp \{-i \alpha \sigma_{3}/2\}\exp \{-i \beta \sigma_{2}/2\} \exp \{i \alpha \sigma_{3}/2\}  \exp \{-i \gamma \sigma_{3}\} \\   &=&   \left(             
  \begin{array}{cc}   
    \cos\frac{\beta}{2} & -\sin\frac{\beta}{2} e^{-i\alpha} \\  
    \sin\frac{\beta}{2} e^{i\alpha} & \cos\frac{\beta}{2}   \\  
  \end{array}
\right)  \left(             
  \begin{array}{cc}   
     e^{-i\gamma} & 0 \\  
0 &   e^{i\gamma} \\  
  \end{array}
\right)   \;.
\end{eqnarray} 
$ u(s) $ is a periodic matrix with $u(0)=u(1)  $, so $ \alpha$, $ \gamma $, and $ \beta $ are periodic functions of $ s $ modulo $2\pi$, $2\pi$, and $ 4\pi $. $ tr(\sigma_{3}u^{-1}\sigma_{3}u)=2 \cos\beta $, 
\begin{equation}
S=\int_{0}^{1}ds\; j \;tr[\sigma_{3} (\frac{1}{2}au^{-1}\sigma_{3}u+iu^{-1}\dot{u}) ]=j\int_{0}^{1}ds\; [a\cos\beta +\dot{\alpha}(\cos\beta-1)]\;.
\end{equation}
The Cartan subgroup $ \exp \{-i \gamma \sigma_{3}\} $ is the stationary group of $ e^{iS} $. The integration should be carried over the coset space $ SU(2)/U(1) $ parametrized by $ \alpha $ and $ \beta $.

Let $ \eta=\cos\beta $, $ -1 \leq \eta \leq 1 $, $-\infty < \alpha < \infty$, consider the path integral 
\begin{equation}
G[\eta(0),\alpha(0);\eta(1),\alpha(1)]=\frac{1}{n_{(j,C)}} \int \prod_{s} D_{j}u(s) \; \exp\{iS\}= \int \prod_{s} D_{j}\eta(s) D_{j}\alpha(s)\; \exp\{iS\}
\end{equation}
with the fixed boundary condition $ \eta(1)=\eta(0)=\eta'$, $ \alpha(1)=\alpha(0)+2\pi n $, $ \alpha (0)=\alpha'$, $ \alpha (1)=\alpha''$, $ n \in \mathbb{Z} $. $ D_{j}\eta D_{j}\alpha = \frac{j}{2\pi}  d\eta d\alpha $. The discretized version of the path integral is 
\begin{equation}
G(\eta',\alpha';\eta',\alpha'') =  \lim_{N\rightarrow \infty} (\frac{j}{2\pi})^{N} \int\prod_{k}  d\eta_{k} d\alpha_{k}\; \exp\{i \sum^{N}_{k=1}j[ \frac{a\eta_{k}}{N} +(\eta_{k}-1)(\alpha_{k}-\alpha_{k-1}   )  ]
\}
\end{equation} 
with $ \eta_{0}=\eta_{N}=\eta' $, $ \alpha_{0}=\alpha'$, $ \alpha_{N}=\alpha''$. The integration over $ \alpha_{k} $ gives $ \delta $-functions, making the integral localized at the paths with $ \eta_{k}=\eta' $ for $ k=1,2,\cdots,N-1 $.
\begin{eqnarray}\label{q23}
\nonumber  G(\eta',\alpha';\eta',\alpha'+2\pi n)   &=&  \frac{j}{2\pi} \exp\{i\;j \eta' (a+\alpha''        -\alpha'   )\exp\{-i\;j  (\alpha''        -\alpha'   )
\} \\&=&\frac{j}{2\pi} \exp\{i\;j \eta' (a+2\pi n   )\exp\{-i\;j  (2\pi n  )
\}\;.
\end{eqnarray}
Equation (\ref{q23}) is exact for finite $ N $, but the $ N\rightarrow \infty $ limit must be taken to make (\ref{a1}) valid. The integration over $ \eta' $ gives 
\begin{eqnarray}
\nonumber   G_{\epsilon}(\alpha';\alpha'+2\pi n)&=&  \int^{1+\epsilon}_{-1-\epsilon}d\eta' \;G(\eta',\alpha';\eta',\alpha'+2\pi n)\\   &=&  \frac{\exp\{i\;j  (a+\epsilon  2\pi n   )
\}-\exp\{-i\;j  (a+\epsilon  2\pi n   )
\}}{2\pi i (a +2\pi n   )
}\;,
\end{eqnarray} 
where the regulator $ \epsilon >0 $ is introduced that should be taken to be $ 0 $ at the end. This amounts to adding a small imaginary part to $ \beta $, i.e. $ \beta \rightarrow \beta +i\sqrt{2\epsilon} $. Since 
\begin{equation}
R^{\pm}=\lim_{\epsilon \rightarrow 0} \sum_{n} \frac{e^{\pm i \epsilon 2\pi n}}{2\pi n + a} =\frac{e^{\pm \frac{i}{2} a}}{2\sin\frac{a}{2}} \;, 
\end{equation}
the further integration over $ \alpha' $ and $ n $ gives 
\begin{eqnarray}
\nonumber  G   &=&\sum_{n}  \int^{2\pi}_{0} d\alpha'\; G(\alpha';\alpha'+2 \pi n)\nonumber\\&=&\int^{2\pi}_{0} d\alpha'\; \sum_{n}\frac{\exp\{i\;j  (a+\epsilon  2\pi n   )
\}-\exp\{-i\;j  (a+\epsilon  2\pi n   )
\}}{2\pi i (a +2\pi n   )
} \\&=&\frac{  \sin[(j+\frac{1}{2})a] 
 }{\sin\frac{a}{2} }=\sum^{j}_{m=-j} e^{i am}\;.
\end{eqnarray}
The right-hand sides of (\ref{a1}) and (\ref{a11}) are indeed equal, for which to be possible, the $ j $ dependent path integral measure $ D_{j} \eta  D_{j} \alpha$ should be $ \frac{j}{2\pi}   d\eta d\alpha $.

\section{The completeness of $ \{  \vert D  \rangle_{(A',A)}|\;\forall A    \} $ in $ \mathcal{H}_{ph}[E(A') ]  $ }\label{BBB}

$ \{\vert LA' \rangle_{ph} |\; \forall  L \in \mathcal{L}\} $ composes the complete basis for $ \mathcal{H}_{ph}[E(A') ]   $, so if 
\begin{equation}
\int DA\;g(LA)\vert D  \rangle_{(A',A)}=\int DU\;\vert ULA' \rangle=\vert LA' \rangle_{ph}\;,
\end{equation}
$  \{  \vert D  \rangle_{(A',A)}|\;\forall A    \}  $ will be another set of complete basis for $  \mathcal{H}_{ph}[E(A') ]   $.

$ \forall  \;L \in \mathcal{L} $, suppose $ L=U_{n}T(C_{n-1})U_{n-1}  \cdots T(C_{1})U_{1} $, then
\begin{eqnarray}
\nonumber g(A)g^{-1}(LA) &=& W^{-1}[U_{n-1} T(C_{n-2}) \cdots T(C_{1})U_{1}  A;C_{n-1}] \cdots   W^{-1}[U_{1}  A;C_{1}] \\&=&  W^{-1}[L_{n-1}  A;C_{n-1}] \cdots  W^{-1}[L_{1}  A;C_{1}] \;,
\end{eqnarray}
where $L_{k}  =U_{k} T(C_{k-1}) \cdots T(C_{1})U_{1}   $. The explicit form of $ W^{-1}[L_{k}  A;C_{k}] $ is
\begin{eqnarray}
W^{-1}[L_{k}  A;C_{k}]&=&\nonumber \exp \{-i \oint_{C_{k}}ds\; 2tr[H(L_{k}  A_{i}L^{-1}_{k}+f_{i}(L_{k} ) )]\dot{x}^{i} \} \\&=&\nonumber   \exp \{-\frac{i}{2 \pi}\int d^{3}x \; tr[A_{i}   L^{-1}_{k} HL_{k}    b^{i}( C_{k})  ] \} \exp \{-\frac{i}{2 \pi}\int d^{3}x \; tr[Hf_{i}(L_{k} )  b^{i}( C_{k})  ] \} \;,\\
\end{eqnarray}
where $ b^{i} $ is given by (\ref{16c}), $ f_{i}(L_{k} ) $ is an $ A $ independent term. So 
\begin{eqnarray}\label{coe}
&&\nonumber g(A)g^{-1}(LA) \\&=&\nonumber  \exp \{-\frac{i}{2 \pi}\int d^{3}x \; tr[A_{i} \sum^{n-1}_{k=1}  L^{-1}_{k} HL_{k}    b^{i}( C_{k})  ] \} \exp \{-\frac{i}{2 \pi}\int d^{3}x \; tr[H \sum^{n-1}_{k=1}  f_{i}(L_{k} )  b^{i}( C_{k})  ] \}  \;.\\
\end{eqnarray}
\begin{equation}
\int DA\; g(A)g^{-1}(LA) =\delta [\sum^{n-1}_{k=1}  L^{-1}_{k} HL_{k}    b^{i}( C_{k}) ] \exp \{-\frac{i}{2 \pi}\int d^{3}x \; tr[H \sum^{n-1}_{k=1}  f_{i}(L_{k} )  b^{i}( C_{k})  ] \}\;.
\end{equation}
$\int DA\; g(A)g^{-1}(LA) \neq 0  $ for $ \sum^{n-1}_{k=1}  L^{-1}_{k} HL_{k}    b^{i}( C_{k})  = 0 $, in which case, $  g(A)g^{-1}(LA) $ is $ A $ independent. This is possible only when $L=U  $, and then 
\begin{equation}
\int DA\; g(A)g^{-1}(LA) =\int DU\;\delta (L-U)\;.
\end{equation}
As a result, 
\begin{eqnarray}\label{com1q}
&&\nonumber\int DA\;g(LA)\vert D  \rangle_{(A',A)}\\&=&\nonumber \int dL'\;\int DA\;g(LA)  g^{-1}(L' A)\vert L' A'  \rangle =\int dL'\;\int DA\;g(A)  g^{-1}(L'L^{-1} A)\vert L' A'  \rangle \\&=&\nonumber   \int DU\;\int dL'\;\delta(L'L^{-1}-U)\vert L' A'  \rangle
=\int DU\;\int dL'   \;\delta(L'-U)\vert L'L A'  \rangle\\&=&
\int DU\;\vert UL A'  \rangle  =\vert LA' \rangle_{ph}\;.
\end{eqnarray}
$  \{  \vert D  \rangle_{(A',A)}|\;\forall A    \}  $ composes the complete basis for $  \mathcal{H}_{ph}[E(A') ]   $.

\end{appendix}

\end{document}